\documentclass{aa}  

\usepackage{graphicx,nicefrac}
\usepackage{txfonts}
\usepackage{hyperref}

\usepackage{multirow}

\def\PGPU{$\varphi-$GPU }

\usepackage{orcidlink}
\let\orcid\orcidlink

\begin{document} 

   \title{Dynamical evolution of the open clusters with different star formation efficiencies and orbital parameters}


   \titlerunning{Evolution of the open clusters with the different SFE models}

   \author{M.~Ishchenko \inst{1,2,3}\orcid{0000-0002-6961-8170} \fnmsep\thanks{Corresponding author: {\tt marina@mao.kiev.ua}}
          \and
          V.~Masliukh\inst{1}\orcid{0009-0004-5363-4944}
          \and
          M.~Hradov \inst{1}\orcid{0009-0005-8360-3492}
          \and
          P.~Berczik \inst{1,2,3,4}\orcid{0000-0003-4176-152X} 
          \and
          B.~Shukirgaliyev \inst{5,6,7}\orcid{0000-0002-4601-7065}
          \and
          C.~Omarov\inst{3}\orcid{0000-0002-1672-894X}
          }

   \institute{
Main Astronomical Observatory, National Academy of Sciences of Ukraine, 27 Akademika Zabolotnoho St., 03143 Kyiv, Ukraine 
   \and
Nicolaus Copernicus Astronomical Centre Polish Academy of Sciences, ul. Bartycka 18, 00-716 Warsaw, Poland
  \and
Fesenkov Astrophysical Institute, 23 Observatory str., 050020 Almaty, Kazakhstan
   \and
Konkoly Observatory, Research Centre for Astronomy and Earth Sciences, HUN-REN CSFK, MTA Centre of Excellence, Konkoly Thege Mikl\'os \'ut 15-17, 1121 Budapest, Hungary
   \and
Heriot-Watt University, Aktobe Campus, 263 Zhubanov Brothers Str, 030000 Aktobe, Kazakhstan 
   \and
K. Zhubanov Aktobe Regional University, 34 A. Moldagulova Ave, 030000 Aktobe, Kazakhstan 
   \and
Energetic Cosmos Laboratory, Nazarbayev University, 53 Kabanbay Batyr Ave, 010000 Astana, Kazakhstan 
      }

   \date{Received ; accepted }

 
  \abstract
   { 
   Open star clusters are dynamic systems whose evolution is critically influenced by initial conditions such as star formation efficiency and orbital parameters. Understanding their dissolution mechanisms provides insight into stellar population dynamical mixing in the Milky Way. 
   } 
   { 
   We aim to investigate the dynamical evolution and dissolution of initially non-virialised open clusters by examining how different global star formation efficiencies and orbital characteristics impact the cluster longevity and structural changes. We followed the evolution of the clusters up to their dissolution time on the basis of our calculations. Finally, we also compare our open cluster dynamical evolutionary models with the observed open clusters in our Galaxy's solar vicinity. 
   }
   { 
   Using high-order direct \textit{N}-body simulations, we modelled cluster evolution across different Galactic orbits, systematically varying initial star formation efficiencies to comprehensively explore dissolution mechanisms.
   }
  { 
  Our simulations reveal that open clusters typically survive approximately ten orbital periods, with cluster lifetime being strongly dependent on global star formation efficiency and only marginally influenced by orbital eccentricity. We estimate gas expulsion timescales of $\sim$0.9 Myr, with initial supernova explosions efficiently removing gaseous components from the cluster. The expected lifetime of the cluster (in units of orbital periods) strongly depends on the cluster global star-formation efficiency and only slightly on the orbital eccentricities of the cluster.  
  }
  { 
  The theoretical models demonstrate a remarkable agreement of the Roche-lobe filling parameter with the recent observed Gaia DR3 cluster catalogues in the solar vicinity. By incorporating a mixed sample of clusters with varying star formation efficiencies, we provide a more nuanced understanding of open cluster evolution in the Galactic disc.
  }


   \maketitle
%

\section{Introduction} \label{sec:intro}
Open clusters (OCs) are fundamental building blocks of galaxies and serve as important tracers of galactic structure and evolution \citep{Lada2003, PZ+2010review, Feigelson2013}. Understanding their formation, dynamical evolution, and survival rates is crucial for interpreting the star formation history and chemical enrichment of the Milky Way \citep{Krumholz+19}. However, the processes governing the evolution of OCs, particularly during their early stages, are still not understood.

The early evolution of OCs is complex due to the interplay of various physical processes. OCs form within giant molecular clouds through the gravitational collapse and fragmentation of gas \citep{mckee_ostriker_2007}. During this process, only a fraction of the gas is converted into stars, while the remainder persists as residual gas in the cluster environment. The star formation efficiency (SFE) -- the fraction of gas converted into stars -- is a critical parameter that significantly influences the ability of clusters to remain bound after the expulsion of the residual gas.

The gas expulsion process, driven by feedback from massive stars through stellar winds, ionizing radiation, and radiation pressure, occurs rapidly compared to the dynamical timescale of the cluster, well before supernova events take place \citep{Krause+20, Lewis+2023}. This rapid loss of binding mass can lead to cluster expansion and potential dissolution via violent relaxation. Accurately modelling this process is challenging due to the complex interplay between stellar dynamics, gas hydrodynamics, and stellar feedback \citep{Krumholz+19, Rahner2017, Rahner2019, Pellegrini2020, Pellegrini2020mn}.

Several studies have investigated this phenomenon using combined hydrodynamics and \textit{N}-body simulations \citep[][and many others]{Wall+2019, Li+2019, Fukushima+Yajima2021, sirius3, Sirius2, Guszejnov2022, EKSTER, BIFROST}. However, these methods are computationally expensive, limiting their applicability to modelling a large number of clusters with diverse initial conditions. Interestingly, \citet{sirius3} noted that simplified initial conditions with instantaneous gas expulsion can be a reasonable approximation when focusing on the long-term evolution of clusters.

Various approaches employing \textit{N}-body simulations with the assumption of instantaneous gas expulsion at the outset have been developed \citep[e.g.][]{BK07, GoodwinBastian2006, Smith+11, Bek+17, Shukirgaliyev2021}. These studies have demonstrated that clusters with low SFEs are more likely to dissolve rapidly, while those with higher SFEs have a greater chance of survival. Nevertheless, the precise relationship between SFE, gas expulsion timescales, and cluster survival remains an active area of research \citep{LeeGoodwin2016, Farias+18, Li+2019, Shukirgaliyev2021}.

Of particular interest is the method introduced by \citet{Bek+17}, which employs more physically motivated initial conditions for the density distribution of stars and gas prior to instantaneous gas expulsion. This approach utilises the local-density-driven clustered star formation model of \citet{PP13} to reconstruct the density profile of the residual gas while the cluster is still embedded within it. The model assumes that star formation occurs within a centrally concentrated gas clump, with a constant efficiency per free-fall time, resulting in a star cluster (SC) with a steeper, more centrally concentrated density profile than the gas. This configuration allows even low-SFE clusters to survive the aftermath of gas expulsion by reducing their susceptibility to the gas potential. In particular, \citet{Bek+17} demonstrated that model star clusters with a Plummer profile could survive gas expulsion with a global SFE as low as 15.

In this work, we investigated the long-term evolution of SC with peaked star formation in the centre,  following instantaneous gas expulsion under various Galactic orbital conditions. We adopted the initial conditions for cluster formation based on \citet{Bek+17}, extending our study to elliptical and non-planar orbits. This extension enables us to explore a broader range of realistic scenarios, providing a more comprehensive understanding of cluster evolution within the Milky Way potential.

Section \ref{sec:models} details our methods and simulations, including the Milky Way potential employed and the orbital configurations of the model clusters. We also describe the initial conditions and the simulations conducted. In Sect. \ref{sec:results}, we present our findings on mass loss in the model clusters. Additionally, we develop an analytical framework to characterise the rate and timescale of gas expulsion in Sect. \ref{sec:gas-blow-up}. Section \ref{sec:compar} compares our results with observations, particularly in terms of the Roche-lobe filling factors of Galactic OCs. Finally, we summarise our results and discuss their broader implications in Sect. \ref{sec:conc}.
\section{Methods and models}\label{sec:models}

For the generation of our SC initial conditions, we employed a multi-step approach comprising several key elements: the Milky Way (MW) rotation curve and tidal field; the calculation of the orbital motion of OCs within the Galaxy’s external tidal field; the generation of non-virial equilibrium initial OC models with varying global star formation efficiencies SFEs; and the combination of OC initial data with a chosen initial mass function (IMF) and physical normalisation of the initial conditions. These steps produce a set of initial OC conditions suitable for direct $N$-body modelling of our stellar systems.

\subsection{Galactic tidal field} \label{subsec:gal-field}

For this study, we adopted a static composite Galactic tidal field to model the external influences on the OCs. Specifically, we used the well-known Miyamoto-Nagai potential \citep{MiyamotoNagai75}, which combines a spherically symmetric Plummer model \citep{Plummer_1911} with an axisymmetric Kuzmin disc \citep{Kuzmin1955}. The complex MW potential is represented as a combination of three independent components: the bulge, disc, and halo. This is written as follows:
\begin{equation}
\centering\Phi_{tot} \equiv  \Phi_{B} + \Phi_{D} + \Phi_{H}     
.\end{equation}
Each of these components can be described by the equation
\begin{equation}\label{eq:phi}
    \Phi_{B,D,H}(R, Z)=-\dfrac{G \cdot M_{B,D,H}}{\sqrt{R^2+(a_{B,D,H}+\sqrt{b_{B,D,H}^2+Z^2})^2}}
,\end{equation}
where $a$ and $b$ are the scale radii of the flattened models for each component. Here, $R$ represents the cylindrical radial distance and $Z$ denotes the axial coordinate.

A similar axisymmetric Galactic potential was used in our previous simulations of the dynamical evolution of MW OCs \citep{Just+09, Kharchenko2009, Ernst2010}. Compared to these earlier works, the mass of the halo has been slightly adjusted to achieve a better fit to the circular velocity at the updated location of the Sun in the Galaxy’s rotation curve.

We present our parameters, including the bulge, disc, and halo masses, along with their characteristic scales for the static potential marked as {\tt FIX}. We compare them with other proposed Milky Way potentials from \citet{Bennett2022} and \citet{Bland-Hawthorn2016} in Table~\ref{tab:pot}.

\begin{table}[htbp!]
\caption{Parameters of static Milky Way potential.}
\centering
\resizebox{0.49\textwidth}{!}{
\begin{tabular}{llcc}
\hline
\hline
\multicolumn{1}{c}{Parameter} & Unit & {\tt FIX} & Milky Way \\
\hline
\hline
Bulge mass, $M_{\rm B}$         & $10^{10}~\rm M_{\odot}$ & 1.4 & $\sim1.4$ \\
Disc mass,  $M_{\rm D}$         & $10^{10}~\rm M_{\odot}$ & 9.0 & 6.788 \\
Halo mass,  $M_{\rm H}$         & $10^{12}~\rm M_{\odot}$ & 0.72 & 1.000 \\

Bulge scale length, $a_{\rm B}$ & 1~kpc                     & 0.0 & $\sim0.0$ \\
Bulge scale height, $b_{\rm B}$ & 1~kpc                     & 0.3 & $\sim1.0$ \\

Disc scale length, $a_{\rm D}$ & 1~kpc                     & 3.3 & 3.410 \\
Disc scale height, $b_{\rm D}$ & 1~kpc                     & 0.3 & 0.320 \\
Halo scale height, $b_{\rm H}$ & 10~kpc                    & 2.5 & 2.770 \\
\hline
\end{tabular}
}
\label{tab:pot}
\end{table} 

\subsection{Orbital motion of OCs}\label{subsec:orb-mot}

\begin{figure}[h!]
\centering
\includegraphics[width=0.99\linewidth]{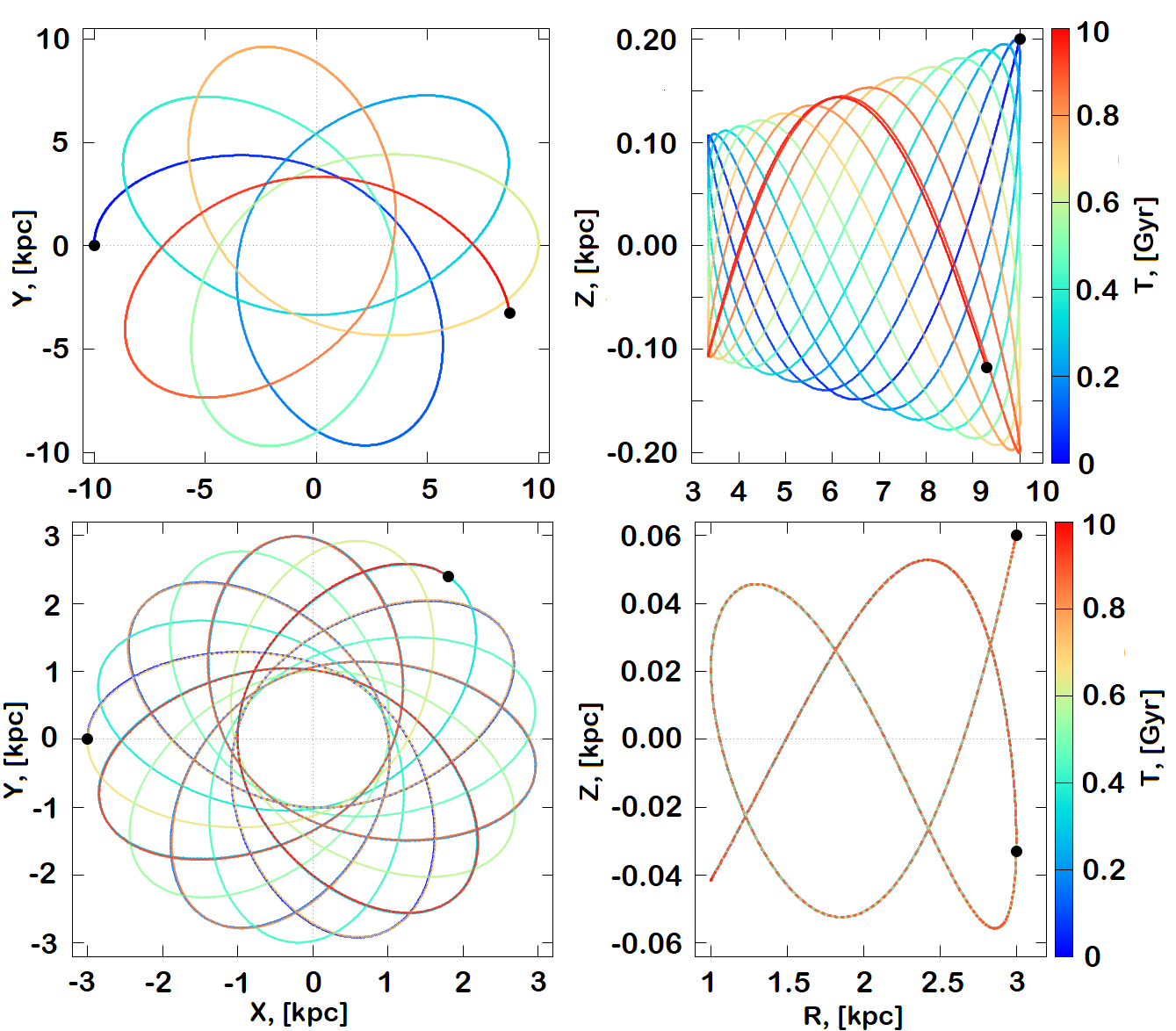}
\caption{Orbital evolution of OC with $e$~=~1/3 is shown in the $X$-$Y$ and $R$-$Z$ planes (where $R$ is the planar galactocentric radius). The integration time is 1 Gyr, as indicated by the colour bar. Upper panel: Orbit with $r_{\text{apo}} = 10\,\text{kpc}$ ($Z=200\,\text{pc}$). Lower panel: Orbit with $r_{\text{apo}} = 3\,\text{kpc}$ ($Z = 60\,\text{pc}$).}
\label{fig:orbits-illu}
\end{figure}

For our set of dynamical simulations, we selected two families of galactocentric orbits with apocentres ($r_\text{apo}$) of 10 and 3 kpc. For these families, we used orbital eccentricities of $e$ = $0$ (circular orbit),  $1/3,$ and $1/2$. We also created models with the same orbital parameters but with an off-plane shift in the initial position, above the Galactic disc plane at  $Z = 200 \, \text{pc}$  and  $Z = 60 \, \text{pc}$  (corresponding to their respective $r_\text{apo}$  values). As an illustration, in Fig. \ref{fig:orbits-illu}, we show two elliptical orbits with  $r_\text{apo} = 10 \, \text{kpc}$  and  $3 \, \text{kpc}$  for  $e$ = 1/3. By using this wide range of orbital parameters, we aim to reproduce a mixture of possible OC orbital configurations in the Galactic disc near the Sun. The set of models with  $r_\text{apo} = 3 \, \text{kpc}$  is added specifically to investigate the potential differences between OCs near the Galactic centre and those in the solar vicinity.

\subsection{Generation of OCs' non-virial initial conditions}\label{subsec:non-vir}

The non-virial initial conditions for open clusters (OCs) prior to gas expulsion, as proposed by \citet{Bek+17}, have been successfully implemented in a series of studies \citep{Bek+18, Bek+19, Shukirgaliyev2021, Bissekenov+2024}. The basic idea of the proposed approach is straightforward: we assume dynamical equilibrium for the self-gravitating two-component system, consisting of a mixture of gas and stars. By assuming a theoretical stellar density distribution and using the approximation of local dynamical equilibrium between self-gravity and the velocity dispersion of the gas-star mixture in each shell, we derive the corresponding equilibrium gas distribution. A key physical assumption in this framework is that star formation occurs locally within each shell on the local free-fall timescale. Using this physical premise, \citet{PP13} developed a semi-analytical model for open cluster formation from centrally concentrated, spherically symmetric gas clumps with a constant SFE $\epsilon_{ff}$ per free-fall time $t_{SF}$. 

\noindent For the stellar-mass density distribution, we used the well-known Plummer model \citep{Plummer_1911}:

\begin{equation}
\rho_{\star}(r) = \frac{3 \cdot M_\star}{4\pi a_{\rm p}^3}\left(1+\frac{r^2}{a_{\rm p}^2}\right)^{-5/2}
,\end{equation}

\noindent where $M_\star$ is the cluster's total stellar mass, and $a_{\rm p}$ is the so-called Plummer radius, which corresponds to the projected half-mass radius of the star cluster.

By varying the SFE per free-fall time and the star formation duration, we created a set of simulations with different global SFEs, specifically 0.15, 0.17, 0.20, 0.25, 0.30, and 0.35. In \citet{Bek+17}, it was shown that a minimal global SFE of 0.15 is necessary to form a bound stellar system after instantaneous gas expulsion from the cluster. Our range of modeled global SFEs is in good agreement with observations of open clusters in the Solar vicinity \citep{Higuchi2009, Murray2011}. 

\begin{figure}[h]
\includegraphics[width=1.0\columnwidth]{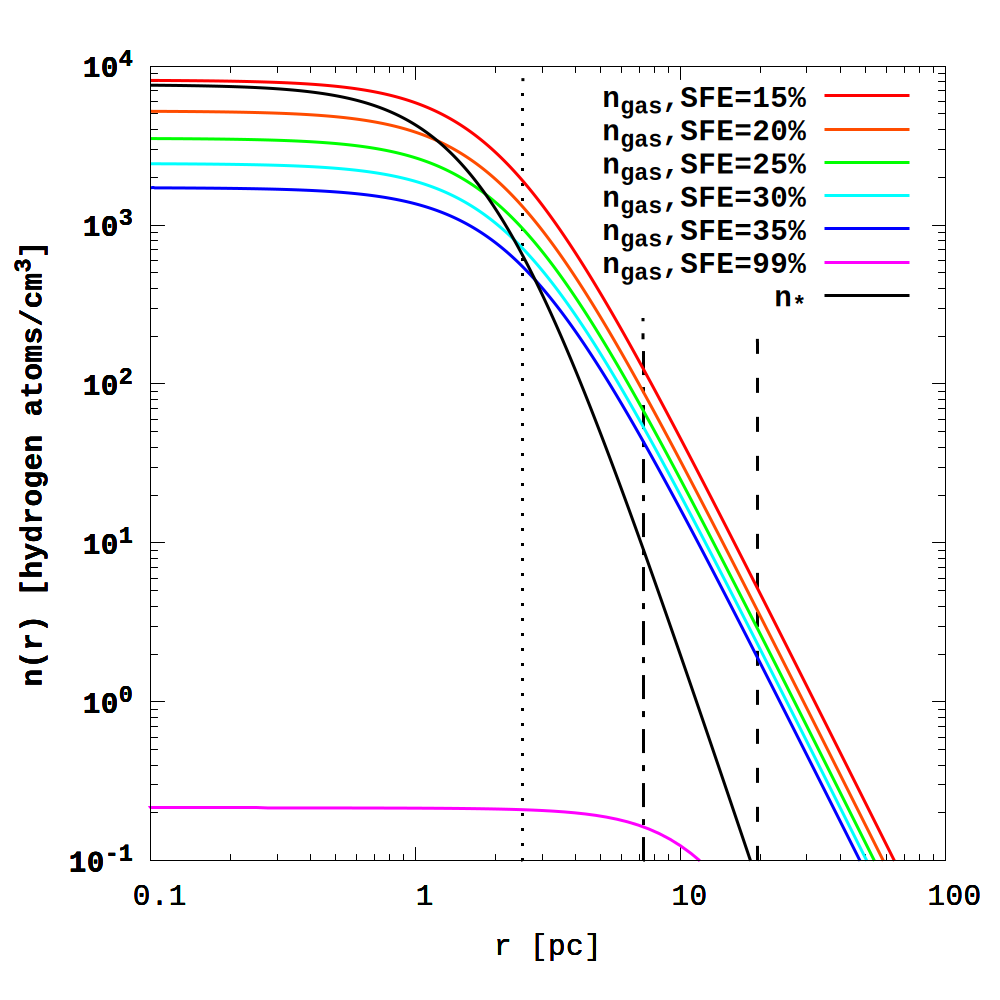}
\caption{
Concentration profiles of cluster's stellar ($n_{\star}$) and gaseous ($n_{gas}$) material are shown for different SFEs, ranging from 0.15 to 0.35 and 0.99. 
The dotted and dash-dotted black vertical lines represent the radii enclosing 50\% and 90\% of the cluster's stellar mass, respectively. The dashed black vertical line marks the simulated cluster boundary, $r_{cluster}$, which is defined as $20 \cdot a_{\rm p}$. 
For the $r_\text{apo} = 10$ kpc and $\lambda_{\rm 0} = 0.09$ models (see Sect.~\ref{subsec:normal}), this corresponds to $\sim$19.5~pc.
}
\label{fig:models-density}
\end{figure}

As an additional set of runs for comparison with previous classical works using initial conditions in virial equilibrium for stellar systems, we also generated a special configuration with a maximum SFE of 0.99. These initial conditions mimic the formation of purely stellar, gas-free OCs (see Fig.~\ref{fig:models-density}).

\subsection{Cluster IMF and physical normalisation}\label{subsec:normal}

The stellar masses are randomly drawn from an adapted Kroupa IMF \citep{Kroupa2001}, with lower and upper mass limits of 0.08 -- 100 M$_{\odot}$. For the stellar metallicity, we adopted the classical solar metallicity value of $Z_{\odot} = 0.02$ \citep{Grevesse1998}. The total stellar mass of our theoretical OC is set to 6000 M$_{\odot}$, representing a typical mass for observed young clusters \citep{Liu2019, Hunt2023}. 

The initial Roche volume filling factor, $\lambda_{\rm {0}}$, is defined as the ratio of the half-mass radius to the Jacobi radius (see definitions of Jacobi mass and Jacobi radius in \cite{Just+09}), where the half-mass radius corresponds to the radius enclosing half the Jacobi mass. Based on earlier studies of OCs \citep{Ernst2010, Ernst2015}, we adopt a standard value of $\lambda_{\rm {0}} = 0.09$ and, in some cases, a value two times smaller, $\lambda_{\rm {0}} = 0.045$. Both values correspond to tidally under-filled clusters.

\begin{figure}[h!]
\centering
\includegraphics[width=0.99\linewidth]{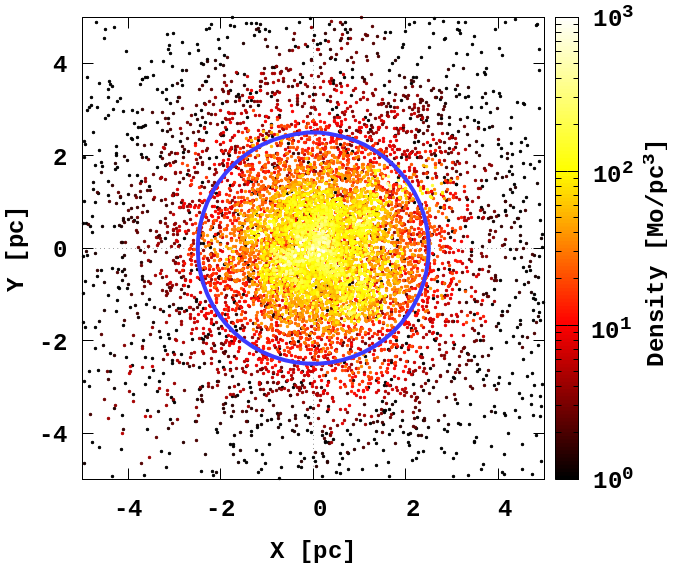}
\caption{
Initial density distribution of OC for $\lambda_{\rm {0}} = 0.09$. The blue circle indicates the half-mass radius of the cluster.
}
\label{fig:ini-sc}
\end{figure}

Using the theoretical OC initial mass and distance from the Galactic centre, we derived the half-mass radius of clusters for the selected $\lambda_{\rm {0}}$. For instance, for $r_\text{apo} = 10$ kpc and a cluster mass of 6000 M$_{\odot}$, the Plummer scale radius is $a_{\rm p} = 1.95$ pc, yielding $r_{\rm hm} = 2.54$ pc for $\lambda_{\rm {0}} = 0.09$ and $a_{\rm p} = 0.97$ pc with $r_{\rm hm} = 1.27$ pc for $\lambda_{\rm {0}} = 0.045$. Figure~\ref{fig:ini-sc} illustrates the initial OC density distribution for $\lambda_{\rm {0}} = 0.09$.

As a final step, we placed our theoretical OC models in an external Galactic field at positions of $r_\text{apo} = 10$ and 3 kpc from the Galactic centre. In addition to the planar models, we ran the full set of models with initial positions at $Z = 200$ and 60 pc above the Galactic plane (see Sect. \ref{subsec:orb-mot}). The complete set of models and values of our SFEs is presented in Table~\ref{tab:init-list}.

For numerical integration, we scaled our dynamical $N$-body modelling using the so-called NB units \citep[also known as H\'{e}non units][]{Henon1971, HeggieMathieu1986}. The chosen NB units are as follows: \\

$M_{\rm NB} = 222.2\; M_{\odot}$,\;\;\;\;\;\;\;\;\;\;\;\;\;\;\;\;\;\; 
$R_{\rm NB} = 1.0$~pc,

$V_{\rm NB} \;= 0.977$~km~s$^{-1}$,\;\;\;\;\;\;\;\;\;\;\;\;
$T_{\rm NB} = 1.0$~Myr.

\begin{table*}[]
\caption{Initial conditions for the set of OC models.}
\begin{center}
\resizebox{0.95\textwidth}{!}{
\begin{tabular}{|c|cccccccccccc|}
\hline
\multirow{3}{*}{\begin{tabular}[c]{@{}c@{}}$M_\star$ = 6000 M$_{\odot}$\\ $N$ = 10 000\end{tabular}} 
& \multicolumn{9}{c|}{\begin{tabular}[c]{@{}c@{}} $\lambda_{\rm {0}}$ = 0.09 \\ $a_{\rm p}$ = 1.95 pc \\ $r_{\rm hm}$ = 2.54 pc \end{tabular}}  
& \multicolumn{3}{c|}{\begin{tabular}[c]{@{}c@{}} $\lambda_{\rm {0}}$ = 0.045 \\ $a_{\rm p}$ = 0.97 pc \\ $r_{\rm hm}$ = 1.27 pc \end{tabular}} \\ \cline{2-13} 
& \multicolumn{1}{c|}{$e$}   & \multicolumn{1}{c|}{$T_{\rm orb}$}   & \multicolumn{7}{c|}{SFE} & \multicolumn{3}{c|}{SFE} \\ \cline{2-13}  & \multicolumn{1}{c|}{}    & \multicolumn{1}{c|}{}    & \multicolumn{1}{l|}{0.15} & \multicolumn{1}{l|}{0.17} & \multicolumn{1}{l|}{0.20} & \multicolumn{1}{l|}{0.25} & \multicolumn{1}{l|}{0.30} & \multicolumn{1}{l|}{0.35} & \multicolumn{1}{l|}{0.99} & \multicolumn{1}{l|}{0.20}    & \multicolumn{1}{l|}{0.25}   & \multicolumn{1}{c|}{0.35}   \\ \hline
\multirow{3}{*}{\begin{tabular}[c]{@{}c@{}}$r_\text{apo}$ = 10 kpc,\\ $Z$ = 0 / 200 pc\end{tabular}} & \multicolumn{1}{c|}{0.0}   & \multicolumn{1}{c|}{270} & +                         & +                         & +                         & +                         & +                         & +                         & +                         & +                            & +                           & +                           \\
 & \multicolumn{1}{c|}{0.(3)} & \multicolumn{1}{c|}{140} & +                         & +                         & +                         & +                         & +                         & +                         & +                         & --                            & --                           & --                           \\
  & \multicolumn{1}{c|}{0.5} & \multicolumn{1}{c|}{130} & +                         & +                         & +                         & +                         & +                         & +                         & +                         & --                            & --                           & --                           \\ \hline
\multirow{3}{*}{\begin{tabular}[c]{@{}c@{}}$r_\text{apo}$ = 3 kpc,\\ $Z$ = 0 / 60 pc\end{tabular}}   & \multicolumn{1}{c|}{0.0}   & \multicolumn{1}{c|}{80}  & +                         & +                         & +                         & +                         & +                         & --                         & +                         & --                            & --                           & --                           \\
 & \multicolumn{1}{c|}{0.(3)} & \multicolumn{1}{c|}{43}  & +                         & +                         & +                         & +                         & +                         & --                         & +                         & --                            & --                           & --                           \\
 & \multicolumn{1}{c|}{0.5} & \multicolumn{1}{c|}{39}  & +                         & +                         & +                         & +                         & +                         & --                         & +                         & --                            & --                           & --                           \\ \hline
\end{tabular}%
}
\end{center}
\tablefoot{
The presented values of $a_{\rm p}$ and $r_{\rm hm}$ are calculated for $r_\text{apo}$ = 10 kpc orbits. The following is true for $r_\text{apo}$ = 3 kpc orbits: 
with $\lambda_{\rm {0}}$ = 0.09, the values are $a_{\rm p}$ = 0.87 pc and $r_{\rm hm}$ = 1.13 pc; with $\lambda_{\rm {0}}$ = 0.045, the values are $a_{\rm p}$ = 0.43 pc and $r_{\rm hm}$ = 0.56 pc.
}
\label{tab:init-list}
\end{table*}

\subsection{Integration procedure}

For the dynamical orbital integration of OCs, including the effects of stellar evolution, we employed the high-order parallel $N$-body code \PGPU\footnote{\PGPU\ $N$-body code: \\~\url{https://github.com/berczik/phi-GPU-mole}}, which is based on a fourth-order Hermite integration scheme with hierarchical individual block time steps (for details, see \cite{Berczik2011} and \cite{Berczik+13}). This code is well tested and has been successfully used to obtain significant results in our prior star-cluster simulations \citep{Bek+17, Bek+18, Shukirgaliyev2021, Ishchenko2024mass-loss}.  

We integrated our OCs up to their dissolution time, with a maximum integration time of 4 Gyr. We define the dissolution of a cluster as the point where the remaining tidal mass decreases to less than 10 \%\ of the initial mass. This criterion is used in conjunction with the central density of the cluster, with dissolution assumed when the central density falls below 0.06~M$_\odot$/pc$^3$ \citep{Gregersen2010}.

\section{Results}\label{sec:results}

\subsection{Mass-loss evolution}\label{subsec:mass-loss}

\begin{figure*}[ht]
\centering
\includegraphics[width=0.99\linewidth]{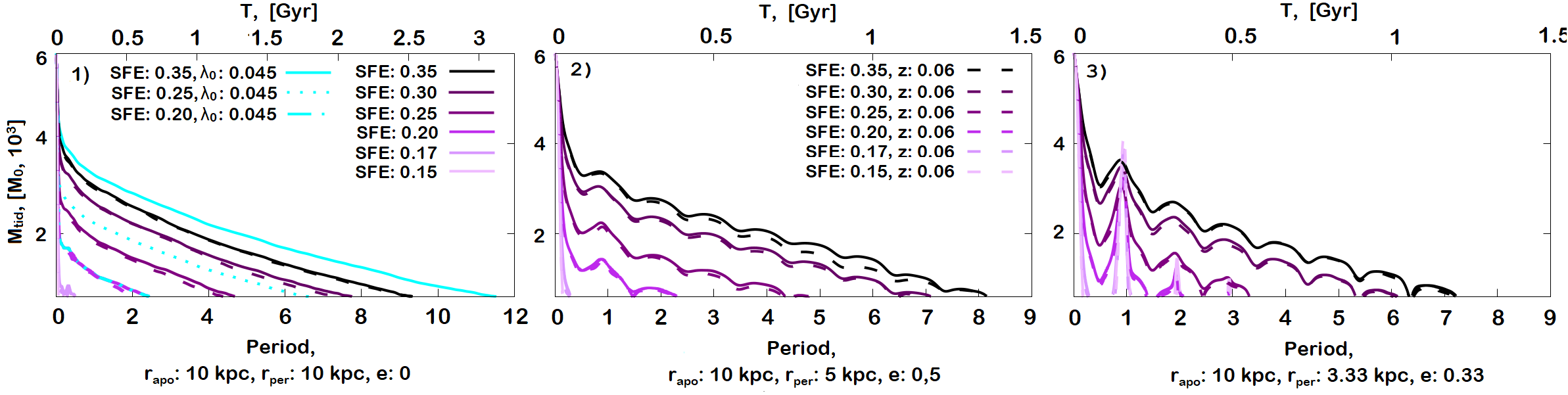}
\includegraphics[width=0.99\linewidth]{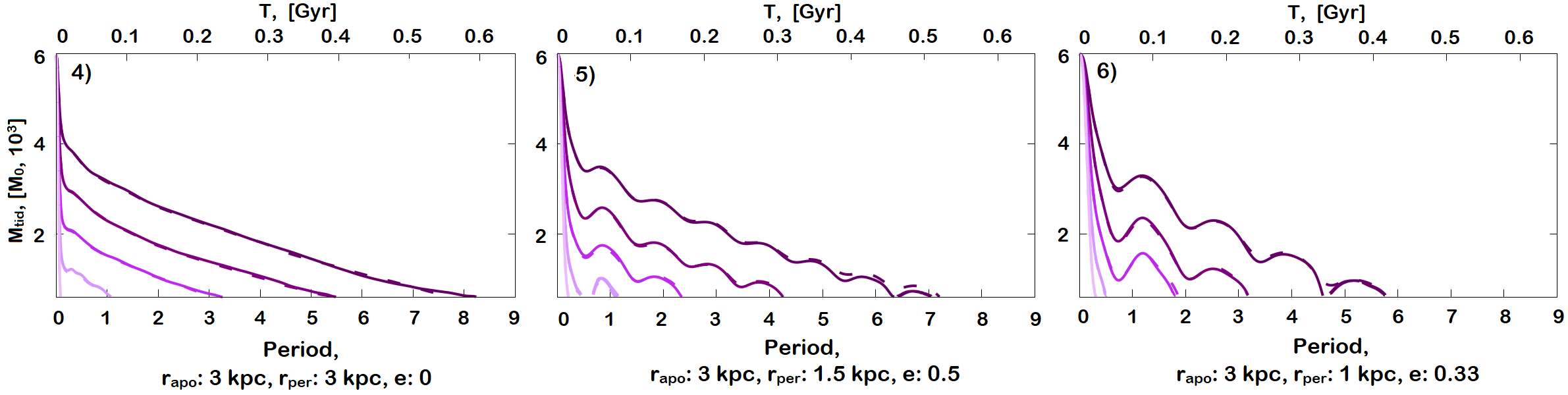}
\caption{Evolution of mass loss in OCs for different SFEs with $\lambda_{\rm {0}}$ = 0.09 (solid lines). Models with $r_\text{apo}$ = 10 kpc and $e$ = 0, 0.33, and 0.5 are shown in the upper panels. Models with $r_\text{apo}$ = 3 kpc are shown in the bottom panels. Dashed lines represent orbits with initial $Z$ values of 0.2 kpc upper panels) and 0.06 kpc (bottom panels) above the Galactic $X-Y$ plane. Cyan lines represent models with $\lambda_{\rm {0}}$ = 0.045 for OCs with $r_\text{apo}$ = 10 kpc.}
\label{fig:mtid}
\end{figure*}

We followed the detailed dynamical and stellar evolution of mass loss in our set of OCs. In Fig. \ref{fig:mtid}, we showed the evolution of the OC tidal mass $M_{\rm tid}$ as a function of time. This figure presents the two main orbital families with $r_\text{apo}$ = 10 and 3 kpc. For each apocentre value, we generated the initial conditions with three different $r_\text{per}$. In Panels 1 and 4, we show the results for circular orbits with $e$ = 0. In Panels 2, 3, 5, and 6, we show the OC models with elliptical orbits of $e$ = 0.33 and 0.5. The colour palette, from black to light purple, represents the initial $\lambda_{\rm {0}}$ = 0.09 values for different SFEs. The solid lines represent the OC models with initial positions exactly in the $X-Y$ Galactic plane. Dashed lines represent the models with initial $Z$ values of 0.2 kpc (Panels 1, 2, 3) and 0.06 kpc (Panels 4, 5, and 6) above the Galactic $X-Y$ plane.

For easier comparison and representation of the OCs with different types of orbits, we present time on the X-axis in units of orbital periods. For example, for circular orbits with $r_\text{apo}$ = 10 kpc, the full period is $\sim$270 Myr, but for orbits with $r_\text{apo}$ = 3 kpc, it is only $\sim$80 Myr. Thus, the real-time evolution of the clusters' mass losses differs significantly. However, by presenting time in orbital periods, we can easily compare the mass losses of clusters on significantly different orbits. The real evolutionary time of the clusters is also shown at the top of each panel.

Comparing all six plots, we can conclude that the models that come close to the Galactic centre have shorter dissolution times. This process is almost independent of their initial SFE values. This can be explained by the influence of the relatively high mass bulge in our {\tt FIX} potential model. OCs with $r_\text{apo}$ = 3 kpc and $e$ = 0.33 and 0.5 are disintegrated after $\sim$3 - 6 rotation periods, as shown in Panels 5 and 6, depending on their initial SFE values. As we can see from Fig. \ref{fig:mtid}, models with low SFE (0.15), regardless of the $r_\text{apo}$ and $e$ values, are destroyed much more quickly -- only after several million years of evolution (less than 0.1–0.2 periods) -- as shown in all panels.

The longest lived models are those with orbits of $r_\text{apo} = 10\ \text{kpc}$ and with $e$ = 0 and 0.33, where we see smooth mass loss for almost all SFEs; this is shown in Panels 1 and 2. We also performed additional simulations with a higher SFE = 0.35, which makes the cluster more concentrated, as shown by the black lines in Panels 1, 2, 3. As expected, in this case the cluster survival rate increases for all orbital eccentricities.

We also ran a few simulations of our basic models (Panel~1) with the reduced initial $\lambda_{\rm {0}}$ = 0.045 parameter. These models are shown in cyan. As we can see, for SFE = 0.20, there is no difference between $\lambda_{\rm {0}}$ = 0.045 and 0.09. However, for SFE = 0.25 and 0.35, the OCs with smaller $\lambda_{\rm {0}}$ demonstrate significantly longer lifetimes.

In Panels 3 and 6, we observe more significant peaks in $M_{\rm tid}$ values for all different SFEs during the first few periods. This behaviour can be understood by considering the tidal radius evolution during the OC’s eccentric orbit. As we know, the OC tidal radius, in a first approximation, is linearly proportional to the OC’s distance from the Galactic centre. So, during the closest approach to the Galactic centre, the OC tidal radius shrinks significantly, and the tidal mass can even approach zero. As the OC flies by the Galactic centre, the tidal radius grows linearly again, and the corresponding tidal mass increases significantly.

In addition to the simple $X-Y$ Galactic planar models, we also ran a set of simulations with the initial position of the cluster above the plane at $Z$ = 0.2 kpc for $r_\text{apo}$ = 10 kpc and at 0.06 kpc for $r_\text{apo}$ = 3 kpc, including all variations of $e$. These models are represented by dashed lines in Fig.~\ref{fig:mtid}. As we can see in the plot, there are no significant differences between these two sets of runs.

In Table~\ref{tab:sc-suvar}, we summarise the lifetime orbital periods of our OCs. As a formal dissolution condition, we set a 10\% limit for $M_{\rm tid}$ in units of the cluster’s initial mass. From the table, we can clearly see that OCs with circular orbits at $r_{\text{apo}}$ = 10 and 3 kpc are the most stable systems. For example, the lifetime of the models with SFE = 0.30 in the first case is $\sim$2.5 Gyr, and in the second case it is $\sim$0.6 Gyr. The special case of OC evolution with an artificially maximum SFE = 0.99 is presented in Fig. \ref{fig:lambda-time} and discussed in Sect. \ref{sec:compar}.
\begin{table}[htbp!]
\caption{Number of periods when OC is considered to be destroyed.}
\centering
\begin{tabular}{lccc|ccc}
\hline
\hline
SFE$^\ast$ & \multicolumn{3}{c}{$r_\text{apo}$ = 10 kpc} & \multicolumn{3}{c}{$r_\text{apo}$ = 3 kpc} \\
    & e 0 & e 0.33 & e 0.5 & e 0 & e 0.33 & e 0.5 \\
\hline
\hline
0.15 & 0.1 & 0.1 & 0.2 & 0.1 & 0.2 & 0.2 \\
0.17 & 0.4 & 0.3 & 0.1 & 1.1 & 0.4 & 0.5 \\
0.20 & 2.2 & 2.3 & 2.0 & 3.2 & 2.3 & 1.8 \\
0.25 & 4.5 & 4.3 & 3.3 & 5.5 & 4.2 & 3.2 \\
0.30 & 7.7 & 7.1 & 6.1 & 8.2 & 7.0 & 5.8 \\
0.35 & 9.3 & -- & -- & -- & -- & -- \\
0.99 & 14 & 12.1 & 10.2 &  12.9 & 12.0 & 10.5 \\
\hline
\end{tabular}
\tablefoot{$\ast$ for $\lambda_{\rm {0}}$ = 0.09.}
\label{tab:sc-suvar}
\end{table} 

\begin{figure}[h!]
\includegraphics[width=1.0\columnwidth]{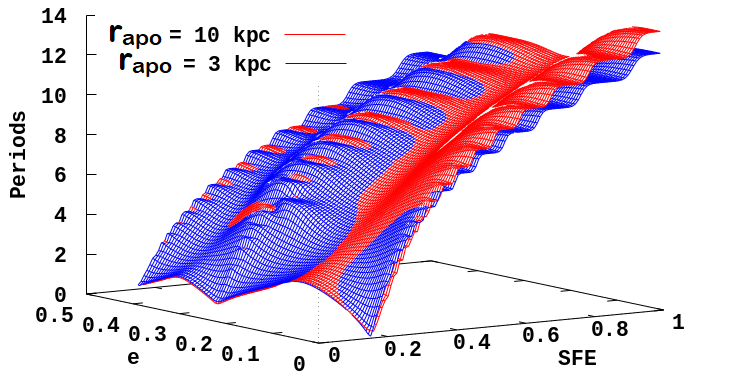}
\caption{Cluster dissolution time in orbital periods as a function of SFE and orbital eccentricity.}
\label{fig:surf}
\end{figure}

In Fig. \ref{fig:surf}, we present a visualisation of the data from Table \ref{tab:sc-suvar} using a simple 3D surface. We observe a clear increasing trend in the cluster dissolution period as a function of SFE. A larger SFE (indicating a smaller remaining gas fraction in the cluster) results in a longer dissolution period. In the figure, the dependence on orbital eccentricity (e) is much less pronounced. The two sets of models with different $r_\text{apo}$ exhibit similar behaviour, with the two surfaces nearly overlapping. In this context, our work complements the study by \citet{Cai2016}, where the authors investigate the evolution of open clusters on Galactic orbits with varying orbital eccentricities. Our results align closely with their conclusions.

\subsection{Evolution of the Roche volume filling factor $\lambda$}\label{subsec:roche}

\begin{figure*}[ht]
\centering
\includegraphics[width=0.99\linewidth]{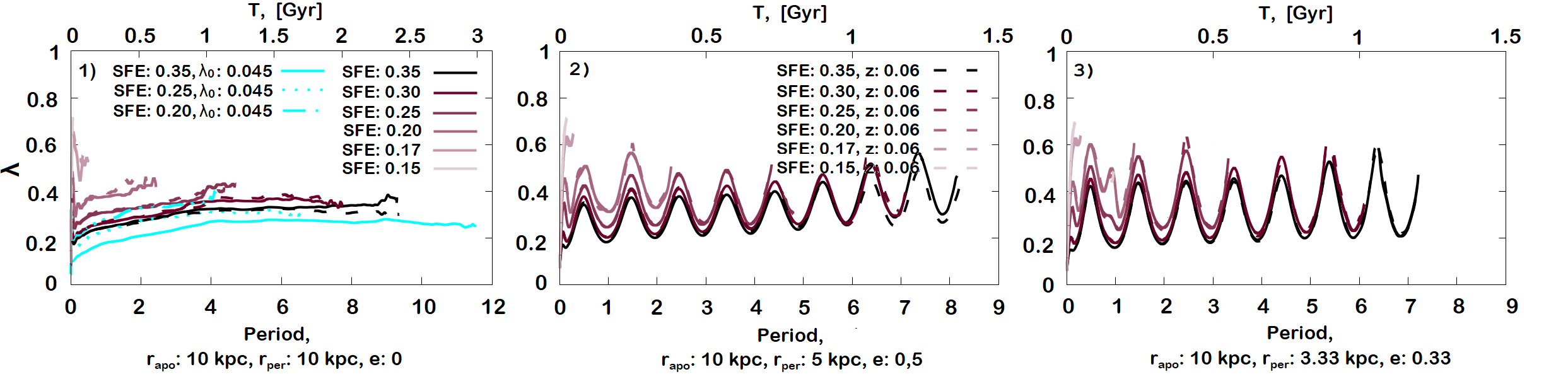}
\includegraphics[width=0.99\linewidth]{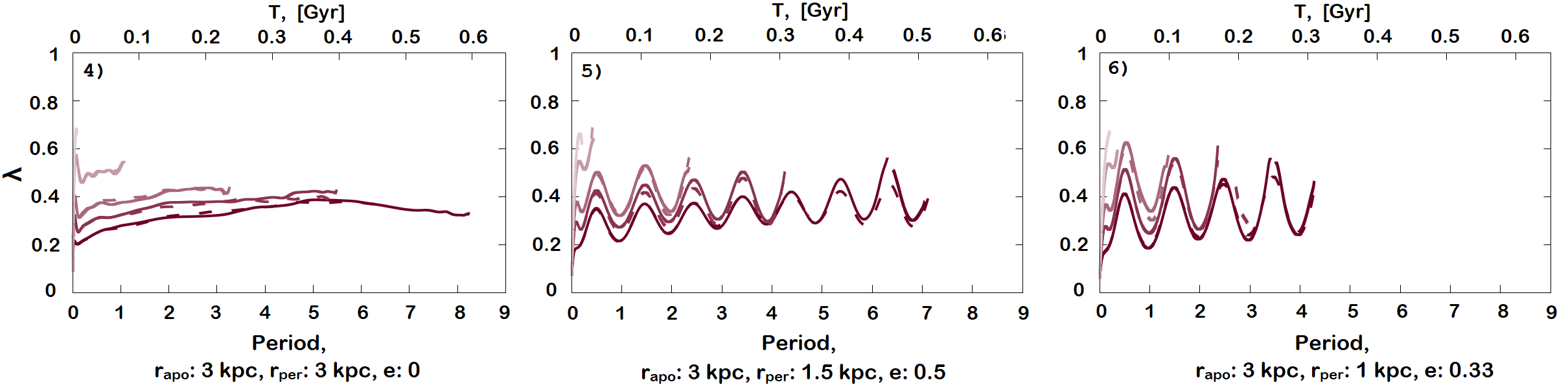}
\caption{Same as in Fig. \ref{fig:mtid}, but for the ${\lambda}$ evolution.}
\label{fig:lambda}
\end{figure*}

In Fig. \ref{fig:lambda}, we show the evolution of the Roche volume filling factor for all our sets of models. Models with circular orbits (Panels 1 and 4) show smoother behaviour in a $\lambda$ range from 0.2 to 0.4 for all SFE values, except for SFE = 0.15. Models with a minimum SFE = 0.15 fall from 0.7 to 0.45 at the beginning of the evolution before the cluster undergoes rapid final dissolution, all panels. Another prominent feature in the panels is the significant oscillation of $\lambda$ as a function of time. The oscillation peaks obviously correspond to the maximum in the $r_\text{apo}$ of the OC positions (e.g. Panel 3 in Fig. \ref{fig:mtid}). The oscillations exhibit a stable period and amplitude.

In Fig.~\ref{fig:lambda} (all panels), we observe no significant differences between planar ($Z$ = 0) and non-planar models. However, we do see notable differences between the two initial $\lambda_{\rm {0}}$ sequences for all different SFEs (see cyan colour in Panel 1).

\begin{figure*}[ht]
\centering
\includegraphics[width=0.97\linewidth]{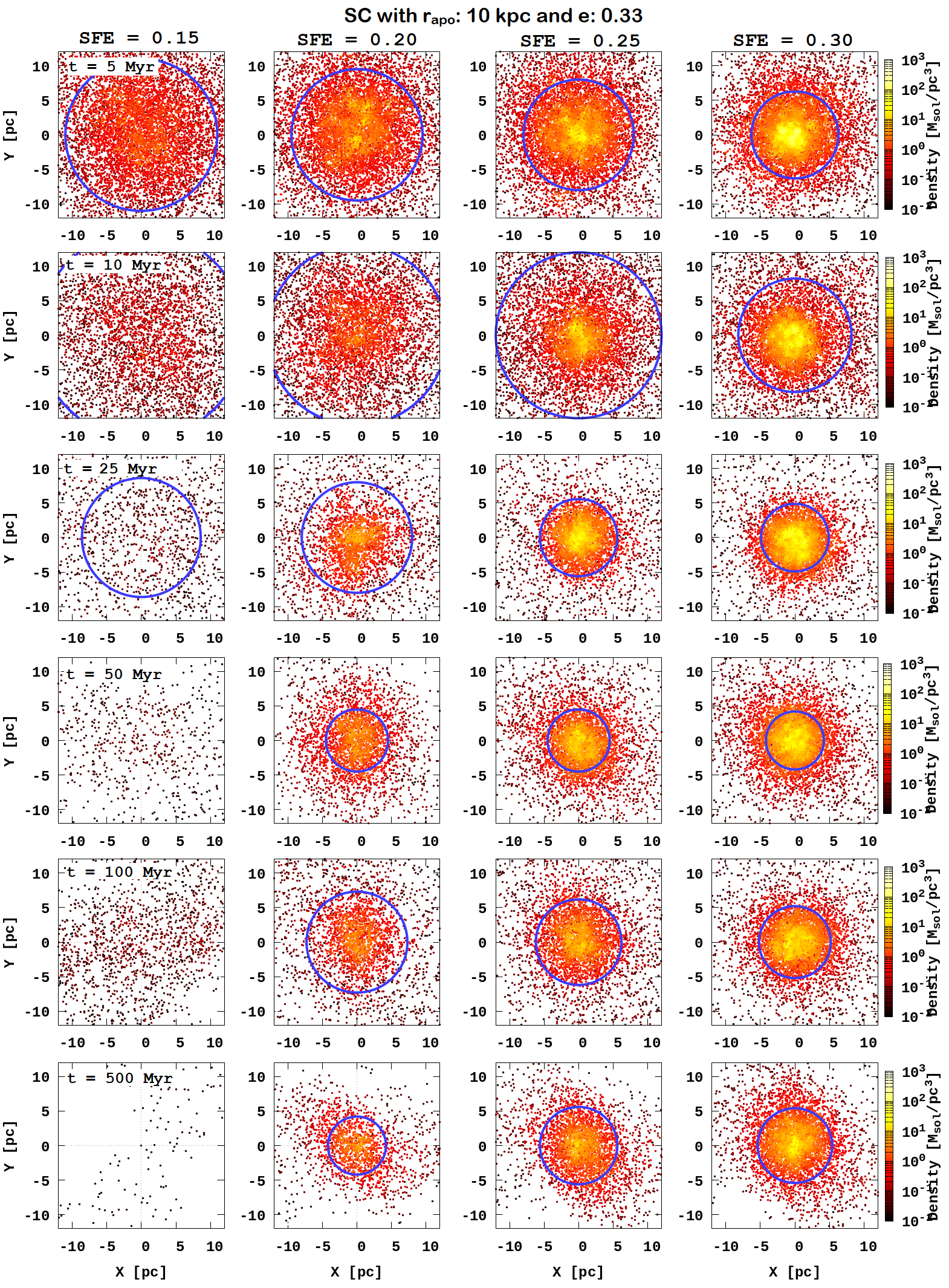}
\caption{Evolution of density distribution in the centre of the OC for an elliptical orbit with $r_\text{apo}$ = 10 kpc and $e$ = 0.33. Snapshots for 5, 10, 25, 50, 100, and 500 Myr are shown. The blue circle represents the half-mass radius.}
\label{fig:pan-10-0.33}
\end{figure*}

In Fig.\ref{fig:pan-10-0.33} and Fig.\ref{fig:pan-3-0.33}, we show the density distribution in the OCs for several models, namely for $r_\text{apo}$ = 10 and 3 kpc with $e$ = 0.33. For each of these models, we also present the particle distributions for different SFEs of 0.15, 0.20, 0.25, and 0.30. The evolution of the half-mass radii (marked as a blue circle) is shown for several time intervals of the OC evolution: 5, 10, 25, 50, 100 Myr, and 500 Myr of the integration. As we can see, the cores for models with SFE = 0.15 and 0.20 were almost dissolved at $T$ = 25 and 50 Myr, respectively. 

\section{Cluster residual star-forming gas expulsion}\label{sec:gas-blow-up}

\begin{figure*}[h]
\centering
\includegraphics[width=0.40\linewidth]{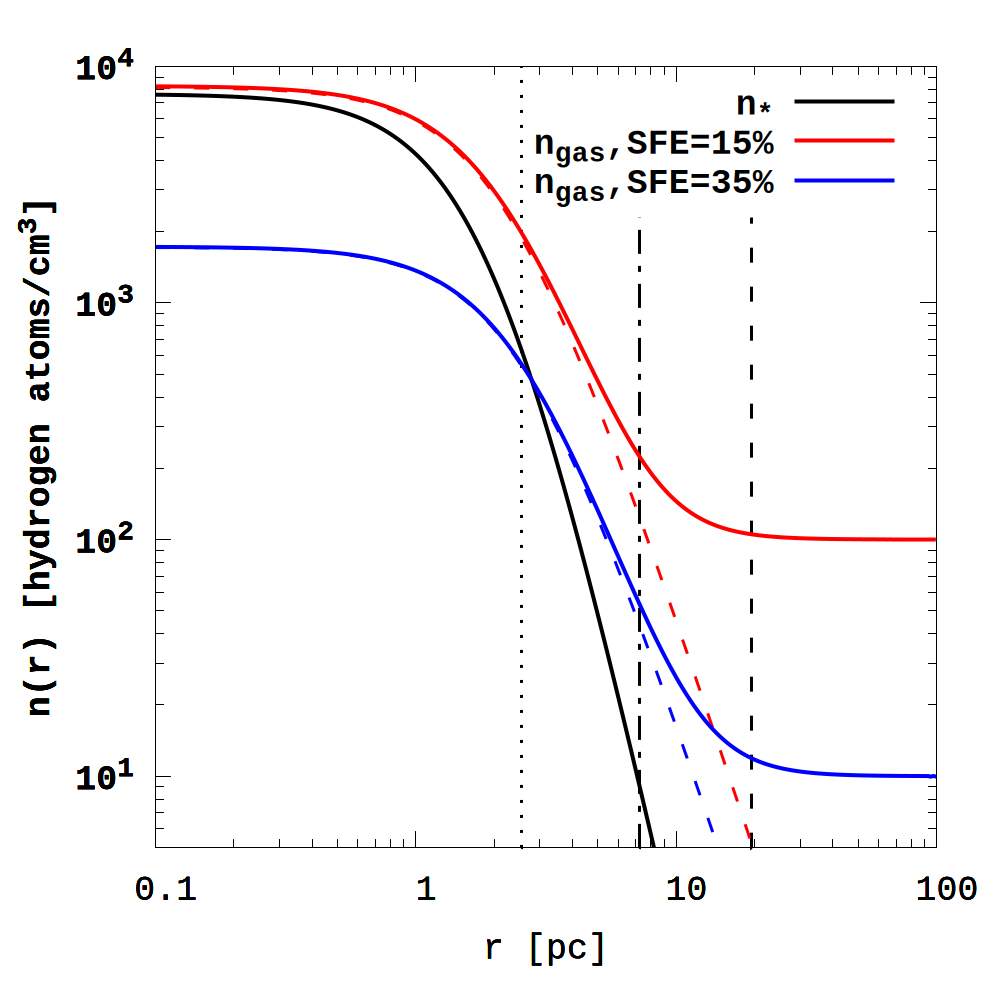}
\includegraphics[width=0.40\linewidth]{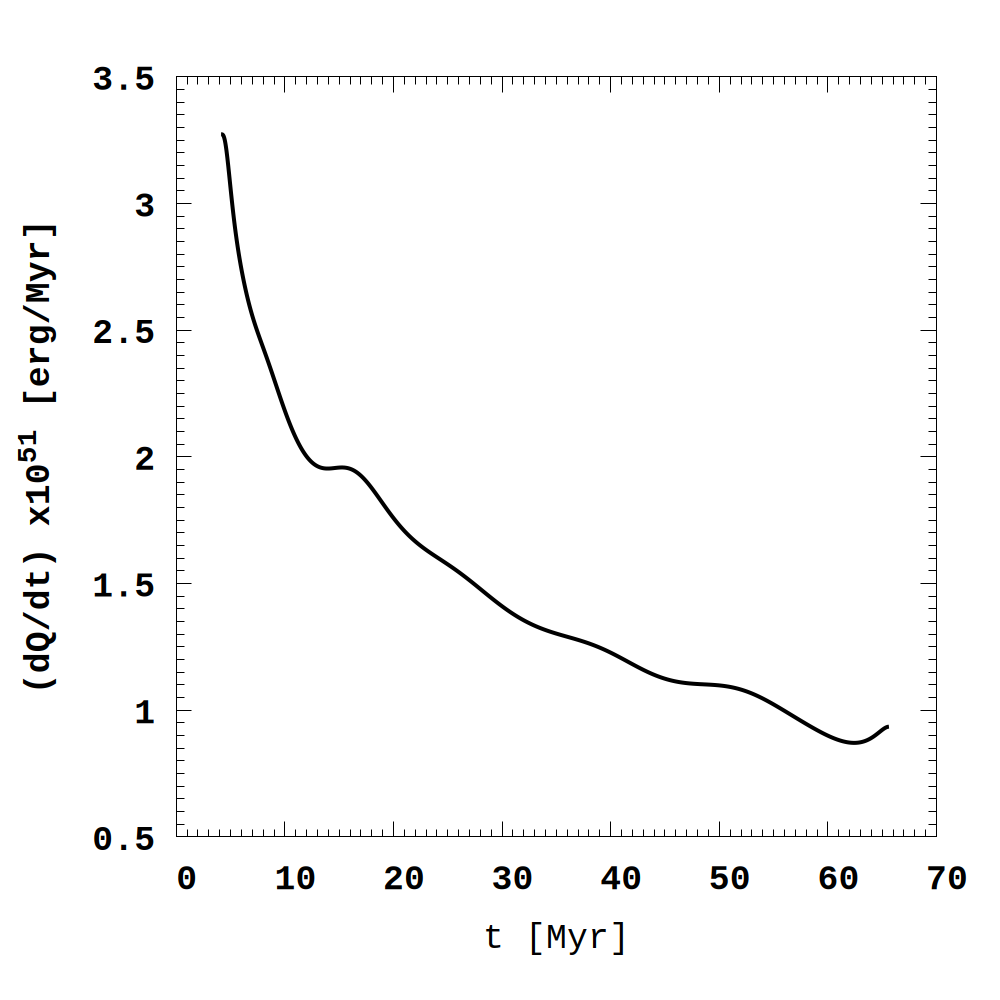}
\includegraphics[width=0.40\linewidth]{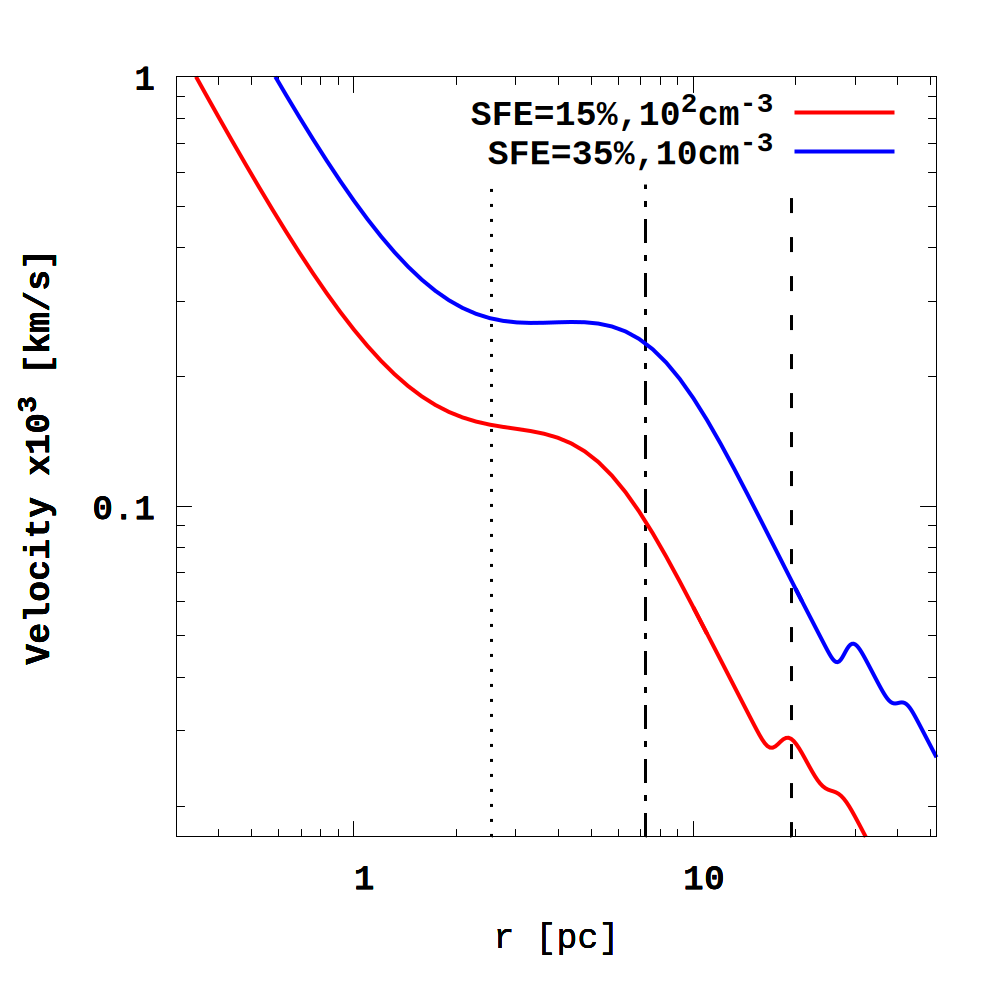}
\includegraphics[width=0.40\linewidth]{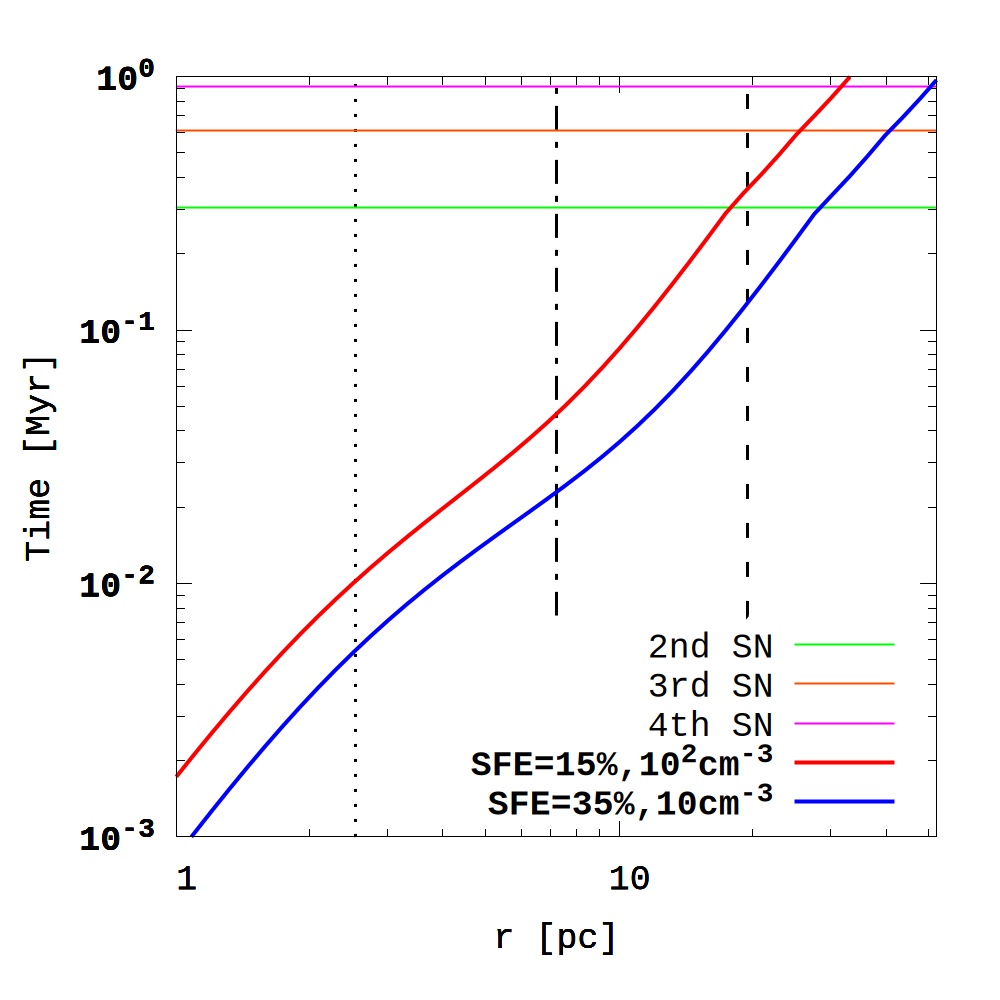}
\caption{Characteristic profiles of gas expulsion processes inside the star cluster (from left to right, top to bottom). Calculated concentration profiles of the cluster stellar $n_{\star}$ and gaseous $n_{gas}$ material for the two cases: SFE = 0.15 with $n_{cloud}$ = 100~cm$^{-3}$ and SFE = 0.35 with $n_{cloud}$ = 10~cm$^{-3}$. The supernova explosion energy rate inside the cluster as a function of time is shown. The shockwave velocity and the shockwave travel time from the centre of the cluster as a function of the front radius. The dotted and dashed-dotted black vertical lines represent the radii within which 50\% and 90\% of the cluster stellar mass are enclosed, respectively. The dashed black vertical line represents the simulated cluster boundary, $r_{cluster}$, which we assume to be equal to 20~$a_{\rm p}$ (for the example model, it is $\sim$19.5~pc).}
\label{fig:gas-blow-up}
\end{figure*}

To estimate the gas expulsion timescale and the residual gas after the supernova explosions in the cluster volume, we studied one example of an open cluster located inside a typical Galactic giant, cold, molecular cloud with a concentration of $n_{cloud} \sim$ 10 - 100 hydrogen atoms per cm$^3$. The cluster stellar matter concentration, $n_{\star}$, is described by the spherically symmetric Plummer model \citep{Plummer_1911}, with a total cluster stellar mass of $M_{\star}$ = 6000 M$_{\odot}$ and a Plummer radius of $a_{\rm p}$ = 1.95 pc. The residual star-forming gas concentration, $n_{gas}$, corresponding to a given global star formation efficiency (SFE), has been recovered using the spherically symmetric semi-analytical model from \cite{PP13}, mentioned in Sect. \ref{sec:models}. The calculated cluster stellar matter $n_{\star}$ concentration profiles for two cases of gaseous external medium, $n_{cloud}$, are presented in \textit{Panel 1} of Fig.~\ref{fig:gas-blow-up}. Here, we show the distributions for two cases: SFE = 0.15 with $n_{cloud}$ = 100 cm$^{-3}$ and SFE = 0.35 with $n_{cloud}$ = 10 cm$^{-3}$.

The young hot stars of the cluster disturb the surrounding gaseous medium by wind-blowing bubbles and by creating extended regions of strong dissociation and ionisation in the remaining gas after star formation \citep{Rahner2017, Rahner2019, Pellegrini2020, Pellegrini2020mn}. For simplicity in further calculations, we do not take into account the existence of wind-blowing bubbles around hot stars. This will result in a certain underestimation of the speed of the cluster gas expulsion process. Instead, we evaluate the phenomenon of hot stars creating the extended HII ionised hydrogen region in the cluster medium. That is, we calculate the radius of the so-called Str\"{o}mgren sphere. For this purpose, we use the equation from Table~5 of the original Str\"{o}mgren paper \citep{Stromgren_1939} (for more details and discussion, see also \cite{Ritzerveld2005}). Using our standard Kroupa IMF \citep{Kroupa2001}, we can simply estimate the number of high-mass ($m \gtrsim$~8~M$_{\odot}$) and high-temperature ($T \gtrsim$~30,000~K) main-sequence stars with O/B spectral type for our model clusters (with total stellar mass $M_{\star}$~= 6000~M$_{\odot}$) as $N_{\rm O/B} \approx$~70. Using this approximate number, we can calculate the ionisation inside the gaseous medium of the cluster. Accordingly, we obtain the Str\"{o}mgren sphere radius equal to $r_{\rm HII}$~= 0.9~pc for the case with SFE~=~0.15 and $r_{\rm HII}$~= 2.3~pc for the case with SFE~=~0.35.

As we can see, a large fraction of the gaseous medium in our model clusters mainly contains non-ionised molecular gas. The strong shockwave from the first supernova stars in the stellar system propagating throughout such an environment produces full dissociation and ionisation of the gas. This results in a noticeably higher compression ratio, more than four times that of a strong shockwave propagating through plasma, at the front of the shockwave moving with a speed of $\sim$100~km/s or less. The shockwave, generated by supernova explosions in the cluster, will greatly slow down when reaching the outside transition region between the gaseous medium of the cluster and the surrounding giant, cold, molecular clouds. After that, its speed may be comparable to $\sim$100~km/s.

Using the stellar evolution library \citep{Kamlah2022}, we calculate that (according to current stellar evolution models) out of the total number of stars ($N_{\star}$ = 10,000 stars) in our model clusters, approximately $N_{\rm SNe} \approx$ 90 stars will explode as supernovae of various types. The start of this process occurs at time $T_{\rm SNe} \approx$ 4.0 Myr after the cluster formation time and stops around 60 Myr. The corresponding chronology of energy transfer by supernova explosions into the gaseous medium of the cluster, calculated under the assumption that each supernova has the same explosion energy $Q_{\rm SN}$ equal to $10^{51}$~erg, is presented in {\it Panel 2} of Fig.~\ref{fig:gas-blow-up}.

The envelope ejected by the supernova explosion, moving through the supernova progenitor surrounding the gaseous medium, generates a strong shockwave. The time evolution of the shockwave's physical parameters, such as shock-front speed, the velocity of gas behind the shock, and the spatial distribution of density and pressure, are described by the well-known Euler equations in fluid dynamics, which are a set of partial differential equations, together with the Rankine–Hugoniot conditions. Each supernova in our model, at the moment of its explosion, is located at some distance from the cluster centre. As a simplifying approximation, we neglect this fact and assume that all supernovae in our model explode near the centre of the cluster. This allows us to simplify the problem from a 3D hydrodynamic problem of a strong explosion in a gaseous medium with non-uniform density distributions to a one-dimensional problem of a strong, spherically symmetric explosion at the centre of a medium with a spherically symmetric non-homogeneous density distribution. Even such a highly idealised problem does not have an exact, complete analytical solution for the general case.
However, this problem does have partial self-similar analytical solutions for the case of a gaseous medium with a power-law density distribution of $\rho_{gas}(r) \propto r^k$, as described, for example, in Chapter~4 of \cite{Sedov1993}. This solution is an extension of the well-known Taylor–von Neumann–Sedov blast wave, which represents a strong explosion inside a uniform gaseous medium.

We adapted the mathematical expression for the cluster residual star-forming gaseous medium density distribution $\rho_{gas}(r)$ to a quasi power-law form as follows (for an arbitrarily chosen $r_0$ and $(r-r_0)^2\ll r_0^2$):

\begin{equation}
\rho_{gas}(r) \simeq \rho_{gas}(r_0) \left( \dfrac{r}{r_0} \right)^{-k(r_0)}
.\end{equation}

In the cases of gaseous medium concentration profiles shown in Panel 1, Fig.\ref{fig:gas-blow-up}, the exponent $k(r)$ increases from a value of zero at the cluster centre to approximately 2.5 at a distance of approximately 5 pc from the cluster centre. It then decreases to approximately 0.5 and lower values, reaching zero at the cluster boundary, which is approximately 20~pc.

According to Eq. 14.4 from paragraph 14 in \cite{Sedov1993}, and taking into account that the cluster gaseous medium density $\rho_{gas}(r) \propto r^k$, the front velocity of a strong spherically symmetric shockwave formed by serial explosions of supernovae at the centre of the cluster can be written as

\begin{equation}
V_{shock}(r)\simeq\dfrac{2}{5-k}\cdot[Q_{\sum\rm SNe}(r)/(\rho_{gas}(r)\cdot r^3)]^{\nicefrac{1}{2}}
,\end{equation}

\noindent where $r$ is the radius of the shockwave front at this moment, $\rho_{gas}(r)$ is the density of the gaseous medium before the front of the shockwave, and $Q_{\sum\rm SNe}(r)$ is the total energy currently deposited by supernovae explosions to the shockwave.

The calculated travel time from the centre of the cluster and the shockwave velocity are presented in {\it Panel 4}, Fig.~\ref{fig:gas-blow-up}. In this figure, the three horizontal coloured lines show the moments when the shockwave, formed by the explosion of the first supernova inside the cluster, is joined by the explosions of the second, third, and fourth supernovae, respectively. These later explosions also begin to deposit energy into the cluster medium.

As seen, at the moment of the explosion of the fourth supernova ($\sim0.9$~Myr after the start of the supernovae explosions), the radius of the front of the total shock wave from the explosions of supernovae significantly exceeds the cluster boundary of $\sim20$~pc.

Since, at this time, the shockwave is moving through the environment of the surrounding molecular cloud, the density distribution behind the front of the shockwave is close to the distribution described by the classical Taylor–von Neumann–Sedov blast-wave solution. Accordingly, almost all the gaseous mass inside the shockwave front is concentrated at a relatively small distance behind the front, towards the centre of the shockwave. Therefore, within the radius that encloses 90\% of the cluster stellar mass, which is almost an order of magnitude smaller than the radius of the shockwave front, only a few percent of the total initial gaseous mass of the cluster remains inside it. In our further dynamical modelling, we neglect the gravitational influence of this remaining gaseous mass on the cluster stars.

In general, our calculations show that the first few explosions of the cluster supernovae are sufficient to produce gas expulsion inside the whole cluster. We can conclude that after $\sim0.9$~Myr from the start of the supernovae explosions, only a few percent of the total initial cluster gaseous mass remains inside the volume.

\section{Comparison with the observational data}\label{sec:compar}

To compare our theoretical OC models with the observed clusters in the solar neighbourhood, we used observations from two different catalogues. The first catalogue, MWSC \citep{Kharchenko2012, Kharchenko2013}, contains the full set of OCs within an  $\sim1.6$~kpc radius around the Sun. It is a heliocentric sample of Galactic clusters constructed using astrometry from the Catalogue of Positions and Proper Motions on the International Celestial Reference Frame \citep{Roeser2010}, in combination with near-infrared photometry from the Two Micron All Sky Survey \citep[2MASS][]{Skrutskie2006}. The MWSC catalogue contains basic parameters for 3061 open clusters \citep{Kharchenko2012}, 202 of which were newly discovered in the MWSC survey \citep{Schmeja2014, Scholz2015}. 

The second, more recent catalogue \citep{Hunt2021, Hunt2023, Hunt2024} contains improved data for all-sky MW OCs based on \textit{Gaia} DR3 observations of more than 729 million sources. The new catalogue contains more than 5647 clusters, with the completeness limit at approximately 1.8 kpc around the Sun. The most important differences between the two catalogues are that the \textit{Gaia} catalogue has a more strongly cut high-quality sample of the 3560 OC objects. These differences are also clearly visible in the range of $\lambda$ parameters of OCs.

\begin{figure*}[ht]
\centering
\includegraphics[width=0.99\linewidth]{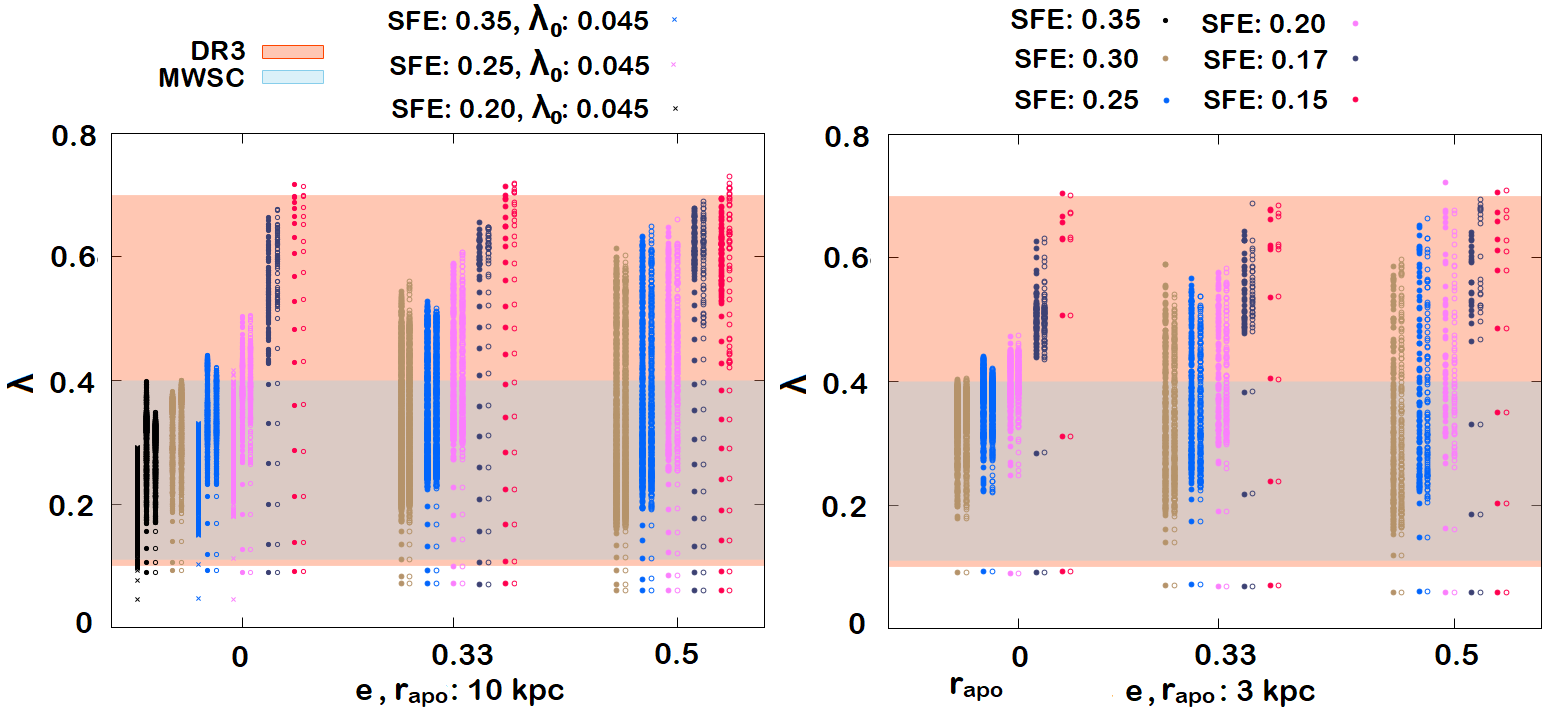}
\caption{Evolution (vertical points) of clusters' Roche-volume filling factors -- $\lambda$ as a function of orbital eccentricities $e$ = 0, 0.33, 0.5 for the two values of orbital apocentres $r_\text{apo}$ = 10 and 3 kpc. With different coloured circles we represent the different values of global SFE. All the basic runs start with the default $\lambda_{\rm 0}$ = 0.09. Additionally, the set of three runs with the two-times smaller $\lambda_{\rm 0}$ = 0.045 presented with the crosses. Filled circles represent the planar orbits ($Z$ = 0), while non-filled circles represent the non-planar orbits. The range of the observed $\lambda$ based on the OC data from \textit{Gaia} DR3 is shown in light red, and the MWSC catalogue is shown in light blue.}
\label{fig:lambda-time}
\end{figure*}

In Fig.~\ref{fig:lambda-time}, we show a comparison of the model $\lambda$ for different $e$ values and at different moments of time (ranging from 0.008 up to 0.7). This shape parameter $\lambda$ increases during the evolution, similarly to the result of \cite{Ernst2015}, and independently of the type of orbit. We observe a clear effect for models with different SFEs. The lower SFE values (0.15 and 0.17) demonstrate faster evolution, quickly growing up to 0.7, which is the upper limit for the observed clusters (\textit{Gaia} DR3). The models with higher SFEs show moderate growth up to 0.4--0.6. Generally, from our set of SC models, we see good coverage of the observed $\lambda$ range in both catalogues. It should be noted that only the recent \textit{Gaia} DR3 catalogue presents OCs with $\lambda$ > 0.4. These observations are in very good agreement with our eccentric orbital models. Models with circular orbits can also reach such high values of $\lambda$, but only in the case of low SFE ($<$0.17).

We also carried out a special set of simulations with the theoretically maximum SFE = 0.99. In this set of simulations, we mimicked the previous (older) gas-free cluster initial conditions. As we can see, these models are compatible with the observations if we assume Galactic eccentric orbits for our clusters.

We also checked the influence of the initial $\lambda_{\rm 0}$ = 0.045 on the $\lambda$ time evolution of the models with circular orbits for $r_\text{apo}$ = 10 kpc (black symbols). The initial $\lambda_{\rm 0}$ has a strong influence on the cluster's filling factor and, as a consequence, on the OC's survival time.

\begin{figure*}[ht]
\centering
\includegraphics[width=0.99\linewidth]{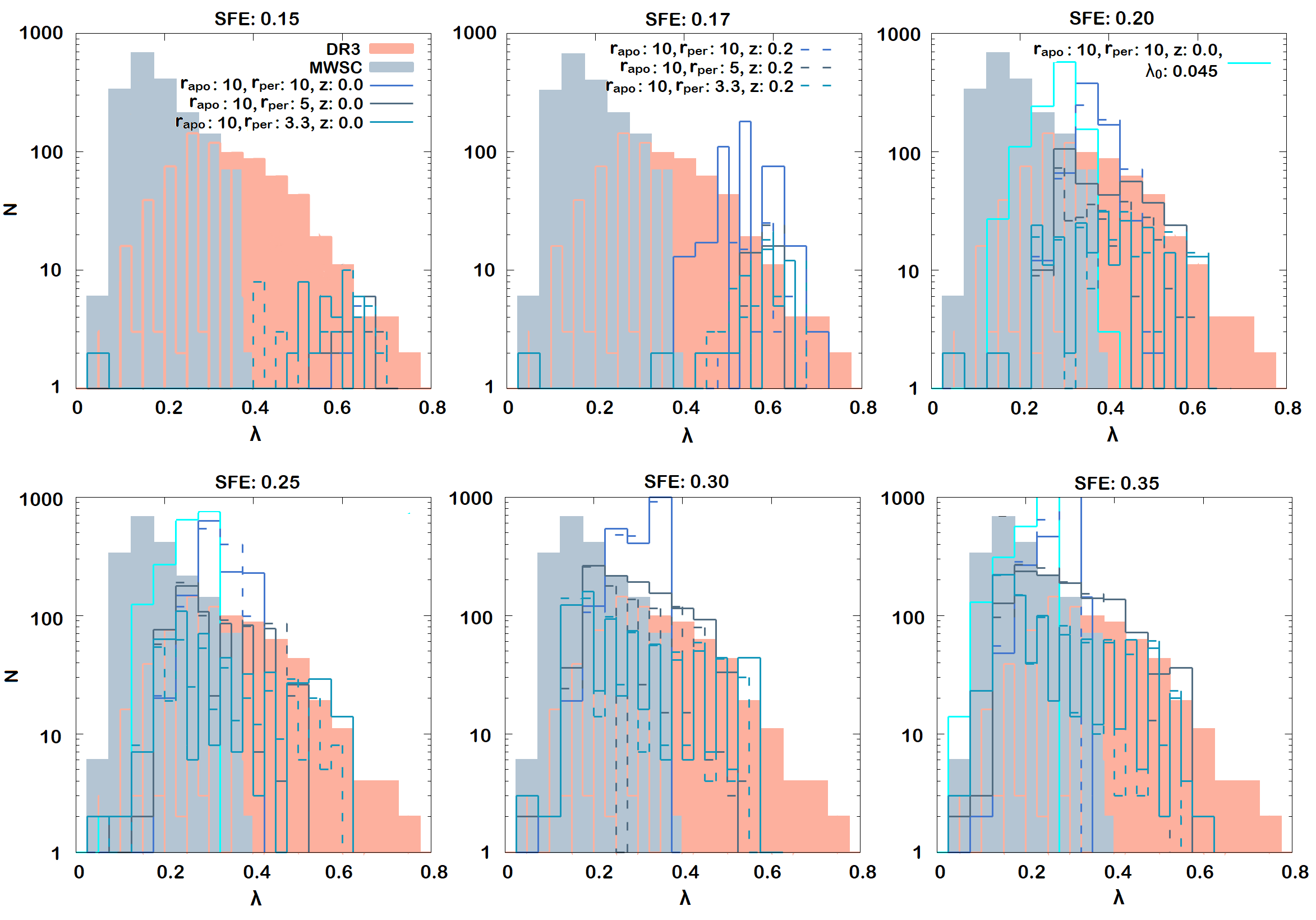}
\caption{Number of clusters (inside the catalogue's completeness limits around the Sun) as a function of the observed cluster's Roche-volume filling factor -- $\lambda$. The distribution of the observational OC data from \textit{Gaia} DR3 is shown in light red, and the MWSC catalogue is shown in light blue. The different panels show the theoretical distributions for different SFE models (with $r_\text{apo}$ = 10 kpc). The lines with different colours show models with different orbital eccentricities. The dotted lines present the results for the non-planar orbits ($Z_0$ = 200 pc). The cyan lines show the results for the three special models with the half-reduced initial $\lambda_{\rm 0}$ = 0.045. }
\label{fig:fix-mtid}
\end{figure*}

In Fig. \ref{fig:fix-mtid}, we present the cluster filling factor distribution for a different set of SFE models, including the special models with SFE = 0.35. Generally, we can conclude that models with lower SFE obviously cannot reproduce the observed cluster distributions. With higher SFE, we managed to reproduce the general trends in both catalogues' distributions. Especially good coverage was achieved for the models with SFE = 0.30 and 0.35.

Our results confirm the general idea that different clusters, at least in the solar vicinity, formed within a given SFE range. Only in this case can we reproduce both the lower $\lambda$ range of the distribution and the higher $\lambda$ range. Small SFE systems can reproduce the higher $\lambda$ range (0.6--0.8), while larger SFEs can reproduce the lower $\lambda$ range (0.01--0.2). Models with non-planar orbits show additional features compared to the planar models. In the end, we observe that the combination of our different model sets can reproduce the wide range of observed distributions. Additionally, it should be mentioned that the number of model clusters in the theoretical curves on our histograms are used directly, without any extra re-normalisation. The good correspondence between observed and theoretical numbers can be interpreted as additional proof of our concept of different SFE formation for OCs in the solar vicinity.

\begin{figure*}[ht]
\centering
\includegraphics[width=0.99\linewidth]{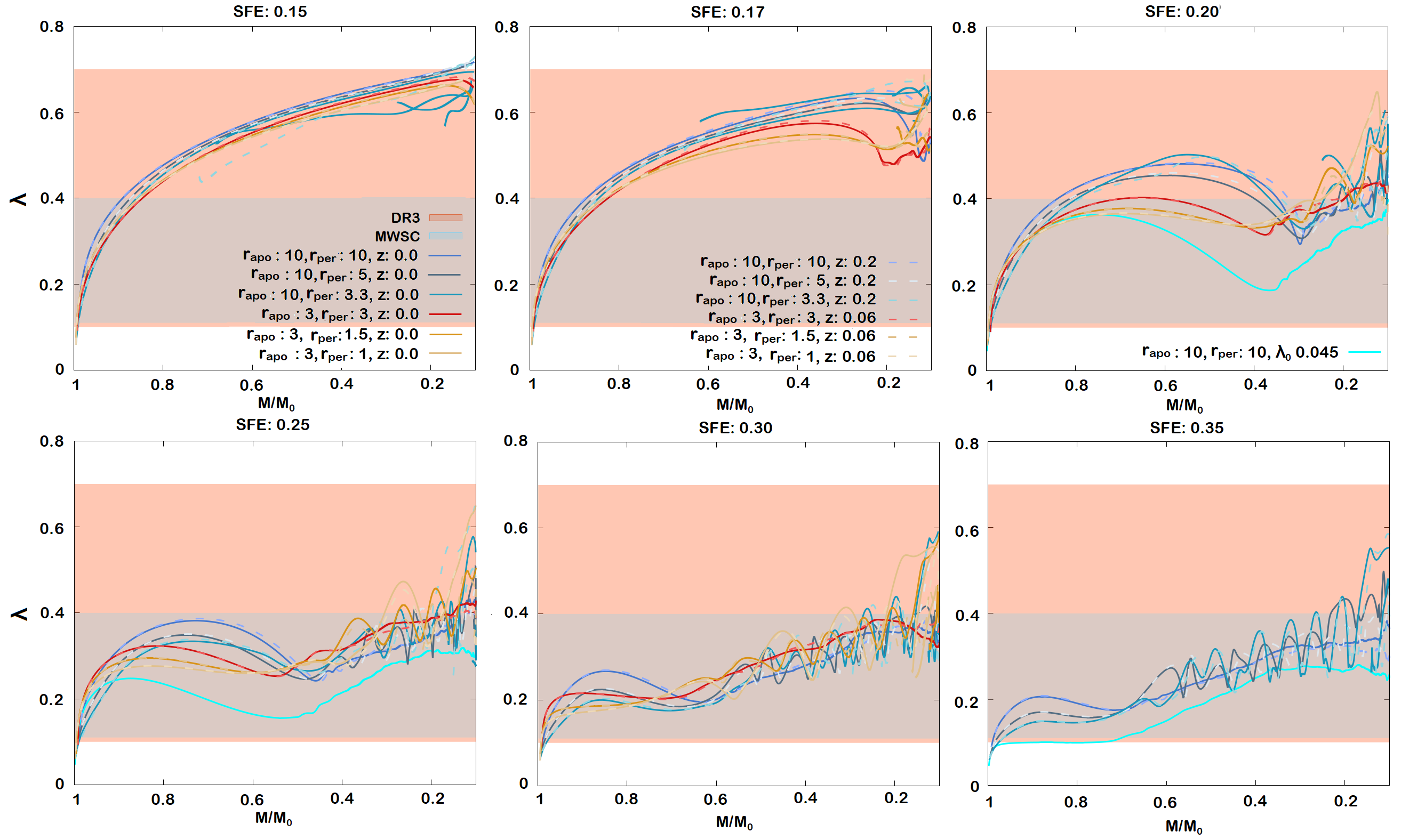}
\caption{Evolution of half-mass Roche-volume filling factor $\lambda$ as a function of the bound mass $M/M_{\odot}$ for SFE = 0.15, 0.17, 0.20, 0.25, 0.30, and 0.35 with $\lambda_{\rm {0}}$ = 0.09. The blue coloured palette curves represent $r_\text{apo}$ = 10 kpc with $e$ = 0, 0.5, and 0.33; while the red coloured palette curves represent the same, but for $r_\text{apo}$ = 3 kpc. Solid curves represent the data for $Z$ = 0, and dashed curves represent $Z$ = 0.2 kpc for $r_\text{apo}$ = 10 kpc and $Z$ = 0.06 kpc for $r_\text{apo}$ = 3 kpc. The cyan curves represent data for $\lambda_{\rm {0}}$ = 0.045 for $r_\text{apo}$ = 10 and $e$ = 0. The distribution of the observational OC data from \textit{Gaia} DR3 is shown in light red, and that from MWSC is shown in light blue.}
\label{fig:gamma-mass}
\end{figure*}

In Fig. \ref{fig:gamma-mass}, we present the $\lambda$ evolution of the OCs as a function of the remaining cluster stellar mass. As we already observed from the previous plots, the lower SFE models describe the whole range of observed Roche-volume filling factors in both catalogues better. The higher SFEs essentially reproduce only the MWSC observed values. We clearly see that the $r_\text{apo}$ = 3 kpc and $r_\text{apo}$ = 10 kpc models are well over-plotted, essentially covering the same $\lambda$ range. As mentioned in the previous paragraph, to describe all the different observed filling factors, we probably need to use a weighted set of different models with varying SFEs (and possibly different initial masses), but such a comprehensive study is well beyond the scope of our current paper.     

\section{Conclusions and discussions}\label{sec:conc}

In this study, we considered the survivability of open clusters with different global star formation efficiency models and Plummer density profiles after instantaneous gas expulsion \citep{Bek+17, Bek+18, Bek+19}. Specifically, we considered model clusters with a fixed initial stellar mass of 6000 M$_{\odot}$ -- corresponding to the Plummer model with $\lambda_{\rm {0}}$ = 0.09 -- with an additional 0.045 in virial equilibrium within the residual gas immediately before the instantaneous gas expulsion.

Our calculations show that a few initial supernova explosions in the cluster are sufficient to expel gas far beyond the cluster boundary ($\approx$ 20~pc). Therefore, within the radius that encloses 90\% of the cluster stellar mass — almost an order of magnitude smaller than the radius of the shockwave front at $\approx$ 0.9~Myr after the onset of supernova explosions — only a few percent of the total initial gaseous mass remains inside the system.

The density profiles of the residual star-forming gas corresponding to a given global SFE were recovered assuming that stars formed with a constant efficiency per free-fall time \citep{PP13, Bek+17}. We then performed direct $N$-body simulations of cluster evolution after instantaneous gas expulsion, assuming they are orbiting in circular and eccentric trajectories in the Galactic disc, with two fixed apocentres at 10 and 3 kpc.

The theoretical models for the Roche-volume filling factor distribution and time evolution coincide well with the observed \textit{Gaia} DR3 and MWSC catalogues. Even our restricted set of models shows good coverage of the clusters' filling parameters. The set of non-planar, orbiting, open-cluster evolution shows a very similar mass loss and filling parameter evolution compared to the planar orbiting star clusters. Compared to previous models \citep{Ernst2010, Ernst2015}, where the authors used 100\% star-forming clusters, our different SFE initial condition models can easily cover the observed cluster filling factor ranges, even up to $\lambda$ = 0.7. To describe all the observed filling factors, we definitely need to use some form of weighted set of theoretical models with different SFEs and initial masses.

As a general conclusion from our set of different SFE models, we observe a relatively short lifespan for typical MW disk OCs that is usually around ten or fewer orbital periods. We note that OCs on eccentric orbits have only slightly shorter lifespans. Our results for the cluster dissolution time from the eccentric orbital simulations show the same behaviour as the models presented in \cite{Cai2016}.

The MW OC dissolution models presented in this paper can serve as a basis for future open-cluster population modelling in the Galaxy. The main feature of our models is the wide range of different star formation efficiency prescriptions. This SFE parameter highlights the strong importance of gas dynamical processes and gas expulsion efficiency for young OCs.
\begin{acknowledgements}
We thank the referee for the helpful comments.

This research has been funded by the Committee of Science of the Ministry of Science and Higher Education of the Republic of Kazakhstan (Grant~No.~BR24992759, AP19677351). 

MI and PB thanks the support from the special program of the Polish Academy of Sciences and the U.S. National Academy of Sciences under the Long-term program to support Ukrainian research teams grant No.~PAN.BFB.S.BWZ.329.022.2023.

PB acknowledges the support by the National Science Foundation of China under grant NSFC~No.~12473017.

BS acknowledges the support provided through the Nazarbayev University Faculty Development Competitive Research Grant Program, with Grant No.~11022021FD2912.

BS acknowledges the use of free version of ChatGPT (OpenAI) for language polishing and manuscript text refinement.
\end{acknowledgements}

\bibliographystyle{aa}
\bibliography{aanda}

\begin{thebibliography}{64}
\expandafter\ifx\csname natexlab\endcsname\relax\def\natexlab#1{#1}\fi

\bibitem[{{Baumgardt} \& {Kroupa}(2007)}]{BK07}
{Baumgardt}, H. \& {Kroupa}, P. 2007, \mnras, 380, 1589

\bibitem[{{Bennett} {et~al.}(2022){Bennett}, {Bovy}, \& {Hunt}}]{Bennett2022}
{Bennett}, M., {Bovy}, J., \& {Hunt}, J. A.~S. 2022, \apj, 927, 131

\bibitem[{{Berczik} {et~al.}(2011){Berczik}, {Nitadori}, {Zhong}, {Spurzem}, {Hamada}, {Wang}, {Berentzen}, {Veles}, \& {Ge}}]{Berczik2011}
{Berczik}, P., {Nitadori}, K., {Zhong}, S., {et~al.} 2011, in International conference on High Performance Computing, HPC-UA 2011, 8--18

\bibitem[{{Berczik} {et~al.}(2013){Berczik}, {Spurzem}, {Wang}, {Zhong}, \& {Huang}}]{Berczik+13}
{Berczik}, P., {Spurzem}, R., {Wang}, L., {Zhong}, S., \& {Huang}, S. 2013, in Third International Conference ''High Performance Computing, 52--59

\bibitem[{{Bissekenov} {et~al.}(2024){Bissekenov}, {Kalambay}, {Abdikamalov}, {Pang}, {Berczik}, \& {Shukirgaliyev}}]{Bissekenov+2024}
{Bissekenov}, A., {Kalambay}, M., {Abdikamalov}, E., {et~al.} 2024, A\&A, 689, A282

\bibitem[{{Bland-Hawthorn} \& {Gerhard}(2016)}]{Bland-Hawthorn2016}
{Bland-Hawthorn}, J. \& {Gerhard}, O. 2016, \araa, 54, 529

\bibitem[{{Cai} {et~al.}(2016){Cai}, {Gieles}, {Heggie}, \& {Varri}}]{Cai2016}
{Cai}, M.~X., {Gieles}, M., {Heggie}, D.~C., \& {Varri}, A.~L. 2016, \mnras, 455, 596

\bibitem[{{Ernst} {et~al.}(2015){Ernst}, {Berczik}, {Just}, \& {Noel}}]{Ernst2015}
{Ernst}, A., {Berczik}, P., {Just}, A., \& {Noel}, T. 2015, Astronomische Nachrichten, 336, 577

\bibitem[{{Ernst} {et~al.}(2010){Ernst}, {Just}, {Berczik}, \& {Petrov}}]{Ernst2010}
{Ernst}, A., {Just}, A., {Berczik}, P., \& {Petrov}, M.~I. 2010, \aap, 524, A62

\bibitem[{{Farias} {et~al.}(2018){Farias}, {Fellhauer}, {Smith}, {Dom{\'\i}nguez}, \& {Dabringhausen}}]{Farias+18}
{Farias}, J.~P., {Fellhauer}, M., {Smith}, R., {Dom{\'\i}nguez}, R., \& {Dabringhausen}, J. 2018, \mnras, 476, 5341

\bibitem[{{Feigelson} {et~al.}(2013){Feigelson}, {Townsley}, {Broos}, {Busk}, {Getman}, {King}, {Kuhn}, {Naylor}, {Povich}, {Baddeley}, {Bate}, {Indebetouw}, {Luhman}, {McCaughrean}, {Pittard}, {Pudritz}, {Sills}, {Song}, \& {Wadsley}}]{Feigelson2013}
{Feigelson}, E.~D., {Townsley}, L.~K., {Broos}, P.~S., {et~al.} 2013, \apjs, 209, 26

\bibitem[{{Fujii} {et~al.}(2021{\natexlab{a}}){Fujii}, {Saitoh}, {Hirai}, \& {Wang}}]{sirius3}
{Fujii}, M.~S., {Saitoh}, T.~R., {Hirai}, Y., \& {Wang}, L. 2021{\natexlab{a}}, \pasj, 73, 1074

\bibitem[{{Fujii} {et~al.}(2021{\natexlab{b}}){Fujii}, {Saitoh}, {Wang}, \& {Hirai}}]{Sirius2}
{Fujii}, M.~S., {Saitoh}, T.~R., {Wang}, L., \& {Hirai}, Y. 2021{\natexlab{b}}, \pasj, 73, 1057

\bibitem[{{Fukushima} \& {Yajima}(2021)}]{Fukushima+Yajima2021}
{Fukushima}, H. \& {Yajima}, H. 2021, \mnras, 506, 5512

\bibitem[{{Goodwin} \& {Bastian}(2006)}]{GoodwinBastian2006}
{Goodwin}, S.~P. \& {Bastian}, N. 2006, \mnras, 373, 752

\bibitem[{{Gregersen}(2010)}]{Gregersen2010}
{Gregersen}, E. 2010, {The Milky Way and beyond} (he Rosen Publishing Group), 35–36

\bibitem[{{Grevesse} \& {Sauval}(1998)}]{Grevesse1998}
{Grevesse}, N. \& {Sauval}, A.~J. 1998, \ssr, 85, 161

\bibitem[{{Guszejnov} {et~al.}(2022){Guszejnov}, {Grudi{\'c}}, {Offner}, {Faucher-Gigu{\`e}re}, {Hopkins}, \& {Rosen}}]{Guszejnov2022}
{Guszejnov}, D., {Grudi{\'c}}, M.~Y., {Offner}, S. S.~R., {et~al.} 2022, \mnras, 515, 4929

\bibitem[{{Heggie} \& {Mathieu}(1986)}]{HeggieMathieu1986}
{Heggie}, D.~C. \& {Mathieu}, R.~D. 1986, in The Use of Supercomputers in Stellar Dynamics, ed. P.~{Hut} \& S.~L.~W. {McMillan}, Vol. 267 (Springer-Verlag), 233

\bibitem[{{H{\'e}non}(1971)}]{Henon1971}
{H{\'e}non}, M.~H. 1971, Ap\&SS, 14, 151

\bibitem[{{Higuchi} {et~al.}(2009){Higuchi}, {Kurono}, {Saito}, \& {Kawabe}}]{Higuchi2009}
{Higuchi}, A.~E., {Kurono}, Y., {Saito}, M., \& {Kawabe}, R. 2009, \apj, 705, 468

\bibitem[{{Hunt} \& {Reffert}(2021)}]{Hunt2021}
{Hunt}, E.~L. \& {Reffert}, S. 2021, \aap, 646, A104

\bibitem[{{Hunt} \& {Reffert}(2023)}]{Hunt2023}
{Hunt}, E.~L. \& {Reffert}, S. 2023, \aap, 673, A114

\bibitem[{{Hunt} \& {Reffert}(2024)}]{Hunt2024}
{Hunt}, E.~L. \& {Reffert}, S. 2024, \aap, 686, A42

\bibitem[{{Ishchenko} {et~al.}(2024){Ishchenko}, {Berczik}, {Panamarev}, {Kuvatova}, {Kalambay}, {Gluchshenko}, {Veles}, {Sobolenko}, {Sobodar}, \& {Omarov}}]{Ishchenko2024mass-loss}
{Ishchenko}, M., {Berczik}, P., {Panamarev}, T., {et~al.} 2024, \aap, 689, A178

\bibitem[{{Just} {et~al.}(2009){Just}, {Berczik}, {Petrov}, \& {Ernst}}]{Just+09}
{Just}, A., {Berczik}, P., {Petrov}, M.~I., \& {Ernst}, A. 2009, \mnras, 392, 969

\bibitem[{{Kamlah} {et~al.}(2022){Kamlah}, {Leveque}, {Spurzem}, {Arca Sedda}, {Askar}, {Banerjee}, {Berczik}, {Giersz}, {Hurley}, {Belloni}, {K{\"u}hmichel}, \& {Wang}}]{Kamlah2022}
{Kamlah}, A.~W.~H., {Leveque}, A., {Spurzem}, R., {et~al.} 2022, \mnras, 511, 4060

\bibitem[{{Kharchenko} {et~al.}(2009){Kharchenko}, {Berczik}, {Petrov}, {Piskunov}, {R{\"o}ser}, {Schilbach}, \& {Scholz}}]{Kharchenko2009}
{Kharchenko}, N.~V., {Berczik}, P., {Petrov}, M.~I., {et~al.} 2009, \aap, 495, 807

\bibitem[{{Kharchenko} {et~al.}(2012){Kharchenko}, {Piskunov}, {Schilbach}, {Röser}, \& {Scholz}}]{Kharchenko2012}
{Kharchenko}, N.~V., {Piskunov}, A.~E., {Schilbach}, E., {Röser}, S., \& {Scholz}, R.-D. 2012, \aap, 543, A156

\bibitem[{{Kharchenko} {et~al.}(2013){Kharchenko}, {Piskunov}, {Schilbach}, {Röser}, \& {Scholz}}]{Kharchenko2013}
{Kharchenko}, N.~V., {Piskunov}, A.~E., {Schilbach}, E., {Röser}, S., \& {Scholz}, R.-D. 2013, \aap, 558, A53

\bibitem[{{Krause} {et~al.}(2020){Krause}, {Offner}, {Charbonnel}, {Gieles}, {Klessen}, {V{\'a}zquez-Semadeni}, {Ballesteros-Paredes}, {Girichidis}, {Kruijssen}, {Ward}, \& {Zinnecker}}]{Krause+20}
{Krause}, M. G.~H., {Offner}, S. S.~R., {Charbonnel}, C., {et~al.} 2020, \ssr, 216, 64

\bibitem[{{Kroupa}(2001)}]{Kroupa2001}
{Kroupa}, P. 2001, \mnras, 322, 231

\bibitem[{{Krumholz} {et~al.}(2019){Krumholz}, {McKee}, \& {Bland-Hawthorn}}]{Krumholz+19}
{Krumholz}, M.~R., {McKee}, C.~F., \& {Bland-Hawthorn}, J. 2019, \araa, 57, 227

\bibitem[{{Kuzmin}(1955)}]{Kuzmin1955}
{Kuzmin}, G.~G. 1955, Publications of the Tartu Astrofizica Observatory, 33, 3

\bibitem[{{Lada} \& {Lada}(2003)}]{Lada2003}
{Lada}, C.~J. \& {Lada}, E.~A. 2003, \araa, 41, 57

\bibitem[{{Lee} \& {Goodwin}(2016)}]{LeeGoodwin2016}
{Lee}, P.~L. \& {Goodwin}, S.~P. 2016, \mnras, 460, 2997

\bibitem[{{Lewis} {et~al.}(2023){Lewis}, {McMillan}, {Mac Low}, {Cournoyer-Cloutier}, {Polak}, {Wilhelm}, {Tran}, {Sills}, {Portegies Zwart}, {Klessen}, \& et~al.}]{Lewis+2023}
{Lewis}, S.~C., {McMillan}, S. L.~W., {Mac Low}, M.-M., {et~al.} 2023, ApJ, 944, 211

\bibitem[{{Li} {et~al.}(2019){Li}, {Vogelsberger}, {Marinacci}, \& {Gnedin}}]{Li+2019}
{Li}, H., {Vogelsberger}, M., {Marinacci}, F., \& {Gnedin}, O.~Y. 2019, \mnras, 487, 364

\bibitem[{{Liu} \& {Pang}(2019)}]{Liu2019}
{Liu}, L. \& {Pang}, X. 2019, \apjs, 245, 32

\bibitem[{{McKee} \& {Ostriker}(2007)}]{mckee_ostriker_2007}
{McKee}, C.~F. \& {Ostriker}, E.~C. 2007, ARA\&A, 45, 565

\bibitem[{{Miyamoto} \& {Nagai}(1975)}]{MiyamotoNagai75}
{Miyamoto}, M. \& {Nagai}, R. 1975, \pasj, 27, 533

\bibitem[{{Murray}(2011)}]{Murray2011}
{Murray}, N. 2011, \apj, 729, 133

\bibitem[{{Parmentier} \& {Pfalzner}(2013)}]{PP13}
{Parmentier}, G. \& {Pfalzner}, S. 2013, \aap, 549, A132

\bibitem[{{Pellegrini} {et~al.}(2020{\natexlab{a}}){Pellegrini}, {Rahner}, {Reissl}, {Glover}, {Klessen}, {Rousseau-Nepton}, \& {Herrera-Camus}}]{Pellegrini2020}
{Pellegrini}, E.~W., {Rahner}, D., {Reissl}, S., {et~al.} 2020{\natexlab{a}}, \mnras, 496, 339

\bibitem[{{Pellegrini} {et~al.}(2020{\natexlab{b}}){Pellegrini}, {Reissl}, {Rahner}, {Klessen}, {Glover}, {Pakmor}, {Herrera-Camus}, \& {Grand}}]{Pellegrini2020mn}
{Pellegrini}, E.~W., {Reissl}, S., {Rahner}, D., {et~al.} 2020{\natexlab{b}}, \mnras, 498, 3193

\bibitem[{{Plummer}(1911)}]{Plummer_1911}
{Plummer}, H.~C. 1911, \mnras, 71, 460

\bibitem[{{Portegies Zwart} {et~al.}(2010){Portegies Zwart}, {McMillan}, \& {Gieles}}]{PZ+2010review}
{Portegies Zwart}, S.~F., {McMillan}, S. L.~W., \& {Gieles}, M. 2010, \araa, 48, 431

\bibitem[{{Rahner} {et~al.}(2017){Rahner}, {Pellegrini}, {Glover}, \& {Klessen}}]{Rahner2017}
{Rahner}, D., {Pellegrini}, E.~W., {Glover}, S. C.~O., \& {Klessen}, R.~S. 2017, \mnras, 470, 4453

\bibitem[{{Rahner} {et~al.}(2019){Rahner}, {Pellegrini}, {Glover}, \& {Klessen}}]{Rahner2019}
{Rahner}, D., {Pellegrini}, E.~W., {Glover}, S. C.~O., \& {Klessen}, R.~S. 2019, \mnras, 483, 2547

\bibitem[{{Rantala} {et~al.}(2023){Rantala}, {Naab}, {Rizzuto}, {Mannerkoski}, {Partmann}, \& {Lautensch{\"u}tz}}]{BIFROST}
{Rantala}, A., {Naab}, T., {Rizzuto}, F.~P., {et~al.} 2023, MNRAS, 522, 5180

\bibitem[{{Rieder} {et~al.}(2022){Rieder}, {Dobbs}, {Bending}, {Liow}, \& {Wurster}}]{EKSTER}
{Rieder}, S., {Dobbs}, C., {Bending}, T., {Liow}, K.~Y., \& {Wurster}, J. 2022, MNRAS, 509, 6155

\bibitem[{{Ritzerveld}(2005)}]{Ritzerveld2005}
{Ritzerveld}, J. 2005, \aap, 439, L23

\bibitem[{{Roeser} {et~al.}(2010){Roeser}, {Demleitner}, \& {Schilbach}}]{Roeser2010}
{Roeser}, S., {Demleitner}, M., \& {Schilbach}, E. 2010, \aj, 139, 2440

\bibitem[{Schmeja {et~al.}({2014})Schmeja, Kharchenko, Piskunov, Roeser, Schilbach, Froebrich, \& Scholz}]{Schmeja2014}
Schmeja, S., Kharchenko, N.~V., Piskunov, A.~E., {et~al.} {2014}, {A\&A}, {568}, A51

\bibitem[{Scholz {et~al.}({2015})Scholz, Kharchenko, Piskunov, Roeser, \& Schilbach}]{Scholz2015}
Scholz, R.~D., Kharchenko, N.~V., Piskunov, A.~E., Roeser, S., \& Schilbach, E. {2015}, {A\&A}, {581}, A39

\bibitem[{{Sedov}(1993)}]{Sedov1993}
{Sedov}, L.~I. 1993, {Similarity and Dimensional Methods in Mechanics} (CRC Press), 496

\bibitem[{{Shukirgaliyev} {et~al.}(2021){Shukirgaliyev}, {Otebay}, {Sobolenko}, {Ishchenko}, {Borodina}, {Panamarev}, {Myrzakul}, {Kalambay}, {Naurzbayeva}, {Abdikamalov}, {Polyachenko}, {Banerjee}, {Berczik}, {Spurzem}, \& {Just}}]{Shukirgaliyev2021}
{Shukirgaliyev}, B., {Otebay}, A., {Sobolenko}, M., {et~al.} 2021, \aap, 654, A53

\bibitem[{{Shukirgaliyev} {et~al.}(2017){Shukirgaliyev}, {Parmentier}, {Berczik}, \& {Just}}]{Bek+17}
{Shukirgaliyev}, B., {Parmentier}, G., {Berczik}, P., \& {Just}, A. 2017, \aap, 605, A119

\bibitem[{{Shukirgaliyev} {et~al.}(2019){Shukirgaliyev}, {Parmentier}, {Berczik}, \& {Just}}]{Bek+19}
{Shukirgaliyev}, B., {Parmentier}, G., {Berczik}, P., \& {Just}, A. 2019, \mnras, 486, 1045

\bibitem[{{Shukirgaliyev} {et~al.}(2018){Shukirgaliyev}, {Parmentier}, {Just}, \& {Berczik}}]{Bek+18}
{Shukirgaliyev}, B., {Parmentier}, G., {Just}, A., \& {Berczik}, P. 2018, \apj, 863, 171

\bibitem[{{Skrutskie} {et~al.}(2006){Skrutskie}, {Cutri}, {Stiening}, {Weinberg}, {Schneider}, {Carpenter}, {Beichman}, {Capps}, {Chester}, {Elias}, {Huchra}, {Liebert}, {Lonsdale}, {Monet}, {Price}, {Seitzer}, {Jarrett}, {Kirkpatrick}, {Gizis}, {Howard}, {Evans}, {Fowler}, {Fullmer}, {Hurt}, {Light}, {Kopan}, {Marsh}, {McCallon}, {Tam}, {Van Dyk}, \& {Wheelock}}]{Skrutskie2006}
{Skrutskie}, M.~F., {Cutri}, R.~M., {Stiening}, R., {et~al.} 2006, \aj, 131, 1163–1183

\bibitem[{{Smith} {et~al.}(2011){Smith}, {Fellhauer}, {Goodwin}, \& {Assmann}}]{Smith+11}
{Smith}, R., {Fellhauer}, M., {Goodwin}, S., \& {Assmann}, P. 2011, \mnras, 414, 3036

\bibitem[{{Str\"{o}mgren}(1939)}]{Stromgren_1939}
{Str\"{o}mgren}, B. 1939, \apj, 89, 526

\bibitem[{{Wall} {et~al.}(2019){Wall}, {McMillan}, {Mac Low}, {Klessen}, \& {Portegies Zwart}}]{Wall+2019}
{Wall}, J.~E., {McMillan}, S. L.~W., {Mac Low}, M.-M., {Klessen}, R.~S., \& {Portegies Zwart}, S. 2019, \apj, 887, 62

\end{thebibliography}

\begin{appendix}
\onecolumn

\section{Evolution of the density distribution in the centre of the OC for model with $r_\text{apo}$ = 3 kpc and $e$ = 0.33}

\begin{figure*}[ht]
\centering
\includegraphics[width=0.88\linewidth]{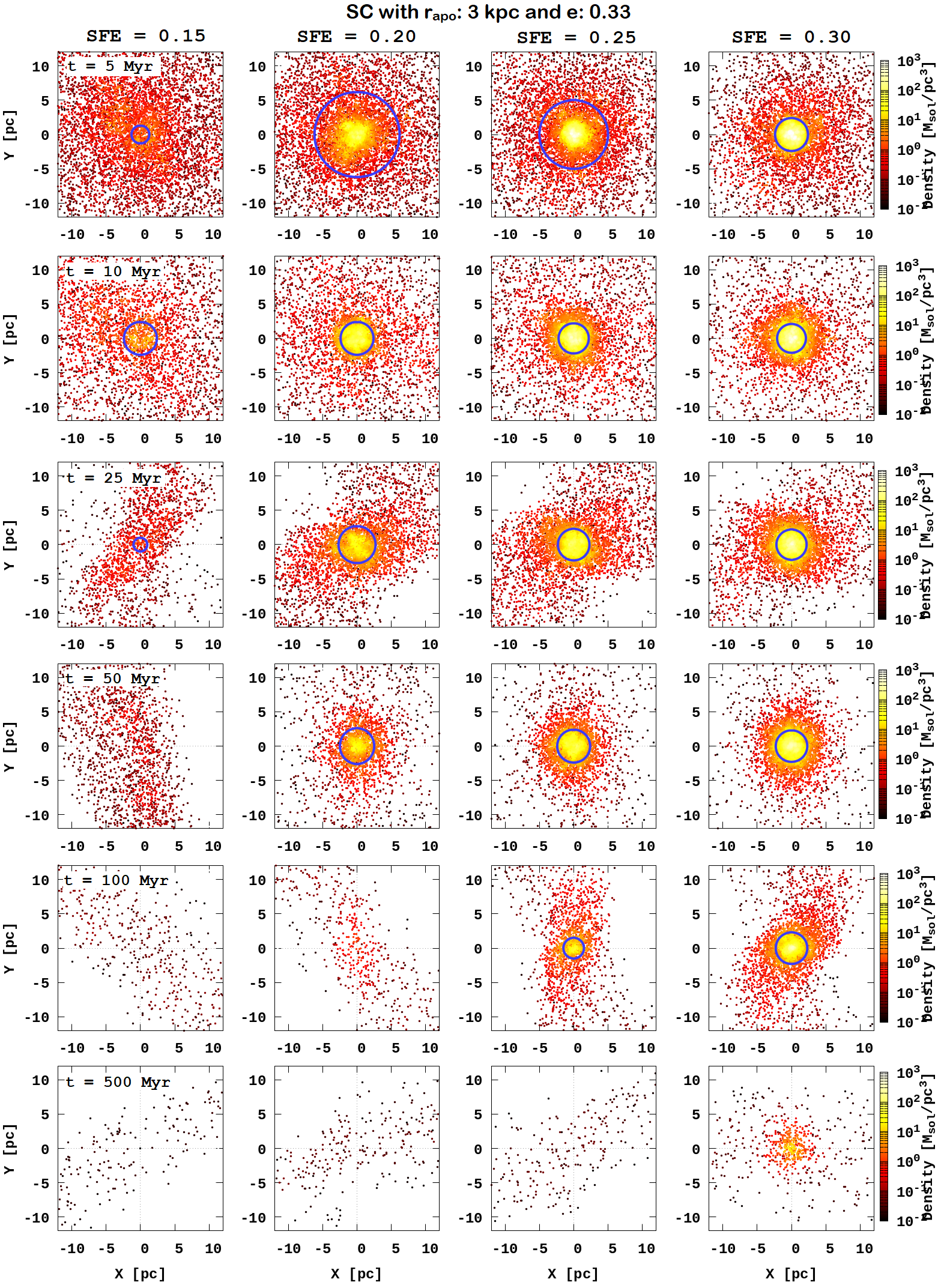}
\caption{Same as Fig. \ref{fig:pan-10-0.33}, but for model with $r_\text{apo}$ = 3 kpc and $e$ = 0.33.}
\label{fig:pan-3-0.33}
\end{figure*}

\end{appendix}

\end{document}